\documentclass[12pt]{article}
\usepackage{times}
\usepackage{graphicx}
\usepackage{cite}

\newcommand{\Rom}[1]{\uppercase\expandafter{\romannumeral #1\relax}}
\newcommand{\Rds}{\widetilde{R}_{xx}}

\usepackage{times}

\usepackage{amsmath}
\usepackage{amssymb}
\usepackage{graphicx}
\usepackage{braket}

\usepackage{xcolor}

\title{Interference of chiral Andreev edge states}

\author
{Lingfei Zhao$^{1\ast}$, Ethan G. Arnault$^{1}$, Alexey Bondarev$^{1}$, Andrew Seredinski$^{1}$,\\
Trevyn F. Q. Larson$^{1}$, Anne W. Draelos$^{1}$, Hengming Li$^2$, Kenji Watanabe$^3$,\\ 
Takashi Taniguchi$^3$, Fran\c cois Amet$^2$, Harold U. Baranger$^{1}$ and Gleb Finkelstein$^{1\ast}$
\\
\normalsize{${}^{1}$Department of Physics, Duke University, Durham, NC 27708, USA}\\
\normalsize{${}^{2}$Department of Physics and Astronomy, Appalachian State University, Boone, NC 28607, USA}\\
\normalsize{${}^{3}$Advanced Materials Laboratory, NIMS, Tsukuba 305-0044, Japan}\\
\\
\normalsize{$^\ast$Corresponding author. E-mail: lz117@duke.edu (L.Z.); gleb@phy.duke.edu (G.F.)}
}

\date{}

\begin{document} 

\baselineskip15pt

\maketitle 

\begin{quote} 
{\bf 
The search for topological excitations such as Majorana fermions has spurred interest in the boundaries between distinct quantum states. Here, we explore an interface between two prototypical phases of electrons with conceptually different ground states: the integer quantum Hall insulator and the s-wave superconductor. We find clear signatures of hybridized electron and hole states similar to chiral Majorana fermions, to which we refer as chiral Andreev edge states (CAES). They propagate along the interface in the direction determined by magnetic field and their interference can turn an incoming electron into an outgoing electron or a hole, depending on the phase accumulated by the CAES along their path. Our results demonstrate that these excitations can propagate and interfere over a significant length, opening future possibilities for their coherent manipulation.}
\end{quote}

The superconducting proximity effect describes the processes in which correlations are induced in a normal metal by a superconducting electrode~\cite{klapwijk_proximity_2004}. The microscopic origin of the proximity effect lies in the Andreev reflections which couple the electron and hole states at the border of a normal metal and a superconductor. In the past few years, interest in Andreev processes has experienced a renaissance, driven by the promise of producing exotic states and excitations, such as Majorana zero modes and chiral Majorana fermions~\cite{BeenakkerRMP2015}, which may be used for topological quantum computing~\cite{Stern_topological_2013,lian_2018_pnas}. Many concrete implementations of Majorana modes have been proposed, relying on superconductors proximitizing either materials with spin-orbit coupling~\cite{lutchyn_majorana_2018} or various quantum Hall systems~\cite{mong_universal_2014,qi_chiral_2010}. 

In this work, we directly probe chiral Andreev edge states (CAES), which result from inducing superconducting correlations in the integer quantum Hall edge states. Semiclassically, CAES result from skipping orbit trajectories, in which an electron turns into a hole and back to an electron upon successive Andreev reflections from a superconductor. Quantum mechanically, this combination of Andreev reflections with the quantum Hall edge states yields fermionic modes in which the electron and hole states are hybridized and propagate chirally along the quantum Hall-superconductor interface~\cite{takagaki_transport_1998,hoppe_andreev_2000,Khaymovich_2010}. Under certain conditions, CAES are predicted to be self-conjugate, becoming chiral Majorana fermions~\cite{Chamon_2010,Tiwari_2013prl,gamayun_two-terminal_2017,chaudhary_vortex_2019}. 

The early search for CAES in \Rom{3}-\Rom{5} semiconductor devices focused on magneto-conductance oscillations in the quantum Hall regime~\cite{eroms_andreev_2005,batov_andreev_2007}. Later, graphene samples in the quantum Hall regime were shown to have enhanced conductance between superconducting contacts~\cite{Komatsu2012,rickhaus_quantum_2012}.
Recent progress in making transparent type \Rom{2} superconducting contacts to both GaAs~\cite{wan_induced_2015} and encapsulated graphene~\cite{BenShalom_quantum_2015,Calado_ballistic_2015} has enabled the observation of several new phenomena, including edge state-mediated supercurrent~\cite{amet_supercurrent_2016,seredinski_full_2019}, crossed Andreev conversion~\cite{lee_inducing_2017}, and inter-Landau-level Andreev reflection~\cite{Sahu_PRL}.

Nevertheless, despite some recent attempts \cite{park_propagation_2017,kozuka_andreev_2018,matsuo_equal-spin_2018}, direct evidence for CAES remains elusive. To conclusively identify the CAES one must demonstrate their propagation along the superconducting contact. Naively, one may expect that any electrical signal spreading along the contact will be shunted by the superconductor. Here, we demonstrate that this is not the case. The main mechanism allowing us to detect the CAES in this work is their interference, which can be described as follows: An incoming electron approaching the superconducting contact is decomposed into a linear combination of CAES~\cite{Khaymovich_2010} propagating along the quantum Hall-superconductor interface with different wavevectors. The accumulated phase difference between these modes can result in the original electron turning into a hole as it exits the opposite end of the interface~\cite{van_ostaay_spin-triplet_2011,lian_edge-state-induced_2016}. The appearance of the hole can be then detected by measuring the voltage on a normal contact located downstream from the grounded superconducting contact~\cite{lee_inducing_2017}. We observe the beating signal which proves that the CAES are formed by a coherent superposition of the electron and hole amplitudes. By further analyzing the interference between the CAES, we show that these modes are on average neutral -- their electron and hole components have a roughly equal weight. These results demonstrate that transport measurements can detect the presence of CAES despite their charge neutrality. Our approach opens the door for detecting the chiral Majorana modes in topological superconductors. 

\begin{figure}
\includegraphics[width=1\textwidth]{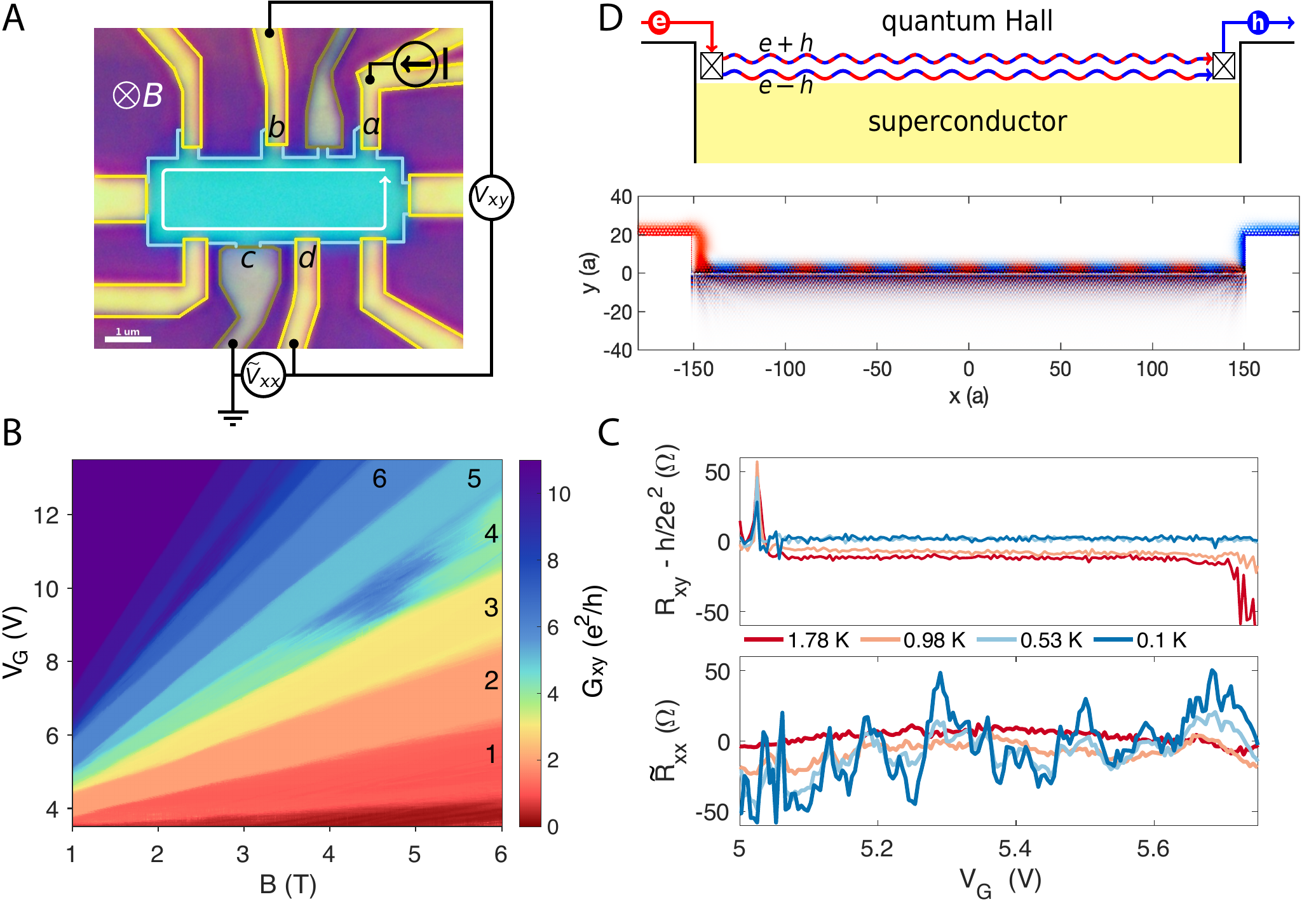}

\noindent {\bf Fig.\,1.} {\bf Andreev reflection in the quantum Hall regime.} (\textbf{A}) Optical image of the sample. Superconducting contacts (gray) are placed between normal contacts (gold). The magnetic field is applied into the plane, resulting in counterclockwise travel of electron and hole when the graphene (blue) is n-doped. We use a four-terminal scheme to measure the Hall voltage $V_{xy}$ simultaneously with the superconductor downstream longitudinal voltage $\widetilde{V}_{xx}$. The sign of $\widetilde{V}_{xx}$ is carefully defined as the voltage of contact \textbf{\textit{d}} minus the voltage at contact \textbf{\textit{c}}. (\textbf{B}) Fan diagram of the zero-bias Hall conductance $G_{xy}$. The filling factors are labeled on the plateaus. Degeneracy of the LLs starts to lift around $B=1$ T, suggesting the high quality of the graphene region. (\textbf{C}) Temperature dependence of zero-bias superconductor downstream longitudinal resistance $\Rds$ measured for $\nu=2$ at $B=3$ T (bottom)and the simultaneously measured Hall resistance $R_{xy}$ (top). To highlight the Hall quantization, $h/2e^2$ is subtracted form $R_{xy}$. The oscillations of $\Rds$ on $\nu=2$ plateau disappears at 1.8 K. (\textbf{D}) At a certain $V_G$, an electron injected into a pair of CAES exits as a hole. The bottom panel shows the electron (red) and hole (blue) wavefunction densities from a tight-binding calculation in the lowest Landau-level.
\end{figure}

\section*{CAES interference}

Our samples are made from graphene encapsulated in hexagonal boron nitride (hBN). The graphene-hBN heterostructure is deposited on a doped Si wafer capped with a 280 nm SiO$_2$ layer, which serves as a back gate. One-dimensional contacts are made to the heterostructure~\cite{wang_one-dimensional_2013} using both normal and superconducting metal electrodes as shown in Fig.\,1A. The superconducting electrodes are sputtered molybdenum-rhenium alloy (MoRe), a type \Rom{2} superconductor with an upper critical field $H_{c2} \approx 10$ T, a critical temperature $T_c \approx 10$ K, and a superconducting gap $\Delta_0 \approx 1.3$ meV. We have previously demonstrated that the interface between MoRe and graphene is highly transparent~\cite{borzenets_ballistic_2016}, which is further confirmed in the Supplementary Fig.\,S3. In addition to the superconducting electrodes, the sample has several normal contacts made of thermally evaporated Cr/Au. 

The lengths of the top and bottom graphene-superconductor interfaces $L$ are 150 nm and 600 nm, respectively. We focus on the bottom contact (\textbf{\textit{c}} in Fig.\,1A). The $L=600$ nm length is chosen so that it falls in the range between the induced superconducting coherence length $\xi_s=\hbar v/\pi \Delta_0 \approx 160$ nm, and the phase coherence length of quantum Hall edge states $\xi_{\varphi}=\hbar v/2\pi k_B T$~\cite{qhcoherence_2008}, which is about $12 \mu$m at $T=0.1$ K. In this estimate, the velocity of the edge states, $v$, is taken to be equal to the Fermi velocity of graphene, $v_F=10^6$ m/s. While this is typically true only for sharp vacuum edges~\cite{edgevelocity_2013}, our simulation (see Supplementary Information section S6) suggests that the CAES velocity is likely lower but comparable to $v_F$, which places $L=600$ nm comfortably between $\xi_{\varphi}$ and $\xi_s$. This condition ensures both that the propagation of the CAES along the contact is quantum-mechanically coherent, and that the crossed Andreev conversion~\cite{lee_inducing_2017} is suppressed. 

Throughout these measurements, we apply a current from the normal contact labeled \textbf{\textit{a}} while keeping the bottom superconducting contact \textbf{\textit{c}} grounded (Fig.\,1A). The current is comprised of a variable DC component, $I$, and a small AC excitation of 10 nA which allows us to measure the differential resistance. To probe CAES propagation along the superconducting contact, we study the longitudinal resistance, $\widetilde{R}_{xx}=d\widetilde{V}_{xx}/dI$, where $\widetilde{V}_{xx}$ is measured between the normal contact \textbf{\textit{d}} and the adjacent grounded superconducting contact \textbf{\textit{c}}. We refer to this quantity as the ``downstream resistance'' to reflect the quantum Hall intuition that the edge states propagate along the chiral direction and thus contact \textbf{\textit{d}} is located downstream from contact \textbf{\textit{c}}. In conventional devices with normal contacts, $\Rds$ would correspond to the longitudinal resistance $R_{xx}$, which equals zero on quantum Hall plateaus and is positive between them (Fig.\,S1).  

We also simultaneously measure the differential Hall conductance, $G_{xy}=dI/dV_{xy}$, where the transverse voltage is measured between the normal contacts \textbf{\textit{b}} and \textbf{\textit{d}} in Fig.\,1A. This quantity remains well-quantized through the relevant range of gate voltages and temperatures below 3 K. This ensures that the bulk of the sample is gapped and that transport occurs only through the edge states. In the data presented, the sample is held at the base temperature below 100 mK unless otherwise specified, and the MoRe remains superconducting at all the magnetic fields ($B<6$ T) and temperatures studied. 

Fig.\,1B shows the map of Hall conductance $G_{xy}$ measured vs. $B$ and gate voltage $V_{G}$. The valley and spin degeneracies in this sample start to lift at $B \approx 1$ T, leading to the appearance of all integer filling factors, $\nu$. The broken symmetry states have smaller activation gaps than the main sequence of filling factors $\nu=4 (n+1/2)$. In order to ensure that the bulk of the sample remains insulating when we vary the current bias and temperature, we focus on the robust filling factor $\nu=2$ (filled lowest Landau level) which has an energy gap on the order of tens of meV. In Fig.\,1C, we plot the simultaneous measurement of $\Rds$ and $R_{xy}$ versus $V_{G}$ in the range corresponding to the Landau level filling factor $\nu=2$ at $B=3$ T. We subtract $h/2e^2$ from $R_{xy}$ to highlight the degree of Hall quantization on the same scale as the variations of $\Rds$. Clearly, the Hall conductance is well-quantized, despite the fact that the current flows through a superconducting drain contact. (The deviations of the plateau level from the quantized value for the individual curves are likely caused by the slow drift of our home-made amplifier.) The observed quantization is in agreement with the Landauer-B\"uttiker formula in Supplementary Information section S3, which shows that $R_{xy}$ measured between the normal contacts \textbf{\textit{b}} and \textbf{\textit{d}} should be quantized regardless of the properties of the drain contact. 

We focus next on the range of gate voltages ($5$ V$<V_G<5.7$ V) in which $R_{xy}$ remains well-quantized at $h/2e^2$ even at higher temperature ($T < 3$ K), ruling out any possible contribution from the bulk. Remarkably, in this range the downstream resistance $\Rds$ shows clear deviations from the zero signal usually expected in the quantum Hall regime (see Supplementary Information section S1). As the temperature is increased, $\Rds$ gradually flattens and approaches zero, and eventually a conventional quantum Hall behavior of zero longitudinal resistance is recovered around $2$ K. Note that this temperature is still very small compared to the quantum Hall gap and indeed $G_{xy}$ remains well-quantized. We have further verified that non-zero $\Rds$ signal is observed only when the grounded contact is superconducting; the resistance measured downstream from a normal contact is strictly zero (see Figs.\,S1\&S2). We conclude that the deviations of $\Rds$ from zero observed for grounded contact \textbf{\textit{c}} is due to superconductivity, whose influence is suppressed by raising the temperature. Incidentally, the vanishing of $\Rds$ observed around $2$ K suggests that contact \textbf{\textit{c}} would be fully transparent in its normal state (see Fig.\,S3 for a discussion of contact transparency in terms of the Blonder-Tinkham-Klapwijk theory). 

Notably, $\Rds$ becomes negative at some gate voltages, suggesting that contact \textbf{\textit{d}} acquires a chemical potential lower than the chemical potential of the grounded contact \textbf{\textit{c}}. 
We attribute this behavior to the following process: An electron approaching the superconductor turns into a linear combination of CAES. For each electron state, a pair of CAES is formed when the proximity effect couples the electron edge state with the hole edge state at the same energy~\cite{Khaymovich_2010} (see Supplementary Information section S2\&S6). Since their wavevectors are different, the two CAES acquire a phase difference while propagating along the superconducting interface, resulting in a beating pattern between the electron and hole components of the wavefunction~\cite{lian_edge-state-induced_2016}. If the CAES interference produces a hole at the end of the graphene-superconductor interface, the hole will flow to contact \textbf{\textit{d}} and lower its chemical potential. Note that in contrast to Ref.~\cite{lee_inducing_2017}, which studies crossed Andreev conversion \emph{across} the superconductor, the negative signal observed here is due to the interference of CAES propagating \emph{along} the contact. As a result, $\Rds$ is sensitive to the phase accumulated along the interface, which makes it dependent on the gate voltage.

In the following, we support our interpretation by conducting tight-binding simulations that illustrate how the CAES propagate along the superconducting contacts, resulting in oscillations of the electron and hole probability along the interface. We then provide further experimental evidence that confirms this picture by showing that the downstream signal is sensitive to the configuration of vortices in the superconducting contacts, and that the measured fluctuations average to zero when sampled over a wide range of magnetic fields.
 
\section*{Tight-binding simulations}

We have conducted detailed tight-binding calculations for a quantum Hall - superconductor interface (see Supplementary Information section S6). The superconducting contacts are modeled by the square lattice; this breaks the valley symmetry at the interface, which would otherwise determine the result of the Andreev reflections through valley isospin conservation~\cite{akhmerov_detection_2007}. We expect the square lattice to provide a generic representation of the rough graphene-MoRe interface. Qualitatively similar results were obtained for a superconductor modeled by a disordered honeycomb lattice. 

Fig.\,1D shows the results of our simulation describing an electron injected in a quantum Hall edge state toward an interface with a superconductor. Note the clear beating pattern between the electron and hole probabilities. The value of the chemical potential is chosen such that the outgoing state is almost purely a hole. We can alternatively obtain an outgoing electron, or any superposition of electron and hole, by changing the chemical potential in our simulation, which corresponds to varying the gate voltage in the experiment. As a result, the calculated probabilities of the outgoing state being an electron, $P_e$, or a hole, $P_h$, show pronounced oscillations (Fig.\,S12). 

Experimentally, the beating pattern between the two CAES is likely to be affected by multiple parameters, such as the interface roughness, disorder potential, electron density profile near the contact, and even positions of vortices in the superconducting contact (Fig.\,S5 and S6). As a result, the downstream resistance measured as a function of the gate voltage acquires a pattern of random but highly reproducible fluctuations (Fig.\,1C), in which the signal is positive or negative depending on whether the superconductor emits predominantly an electron or a hole. We next provide further experimental evidence that supports our interpretation of the non-zero downstream resistance.

\begin{figure}
\includegraphics[width=1\textwidth]{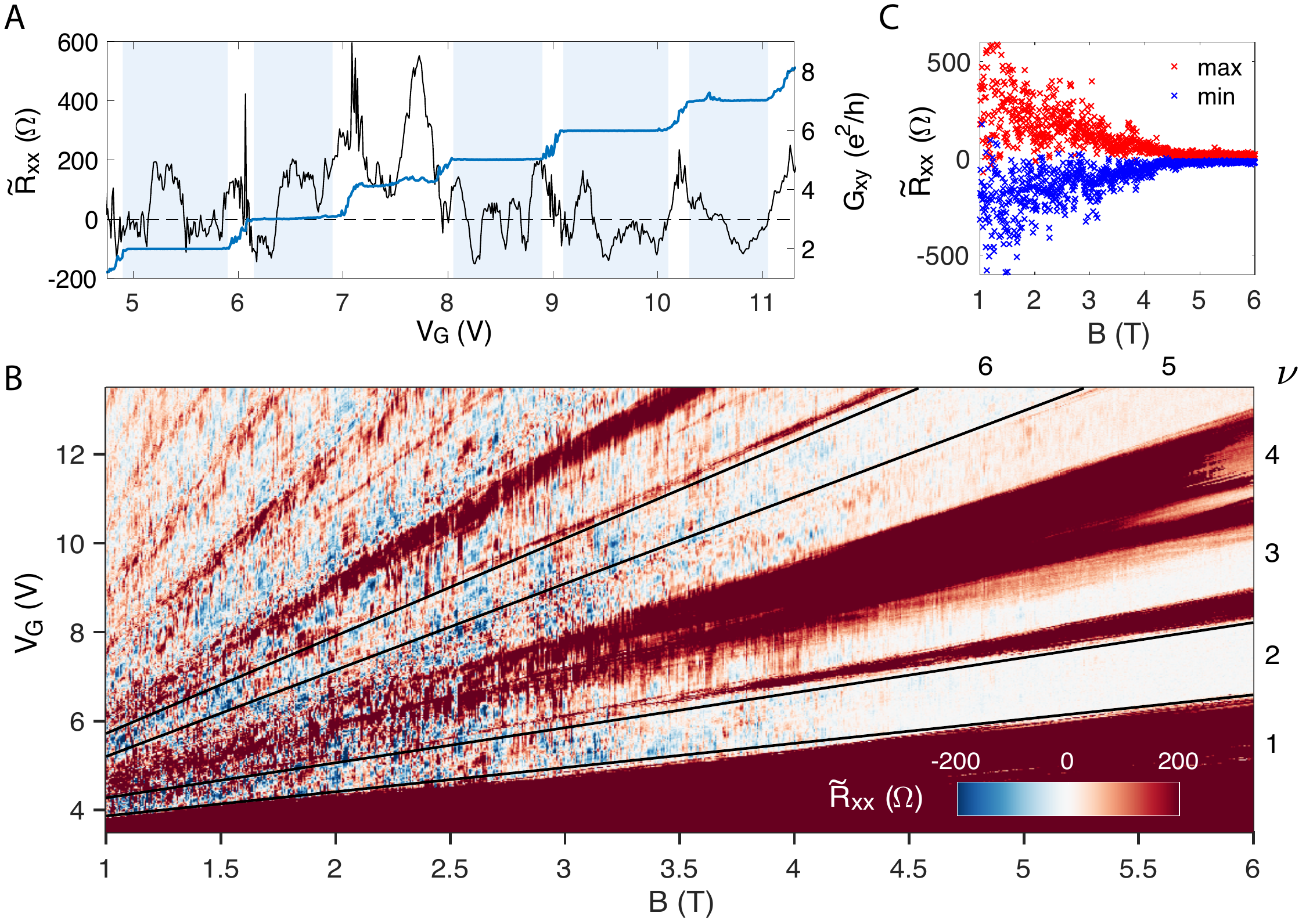}

\noindent {\bf Fig.\,2.} {\bf The interference of CAES on various quantum Hall plateaus and its magnetic field dependence.} (\textbf{A}) Zero-bias $\Rds$ plotted versus gate voltage $V_G$ at $B=3$ T together with the simultaneously measured $G_{xy}$. $\Rds$ oscillates around zero on the well quantized plateaus of $\nu=2, 3, 5, 6, 7$ (shaded regions). (\textbf{B}) Zero-bias $\Rds$ plotted versus gate voltage $V_G$ and magnetic field $B$. Filling factors are labeled based on the $G_{xy}$ fan diagram in Fig.\,1B, with the  boundaries for $\nu=2$ and $6$ drawn as black lines. The oscillations in the well-quantized region gradually die out with increasing magnetic field. (\textbf{C}) Peak values of $\Rds$. The maximum and minimum of $\Rds(V_G)$ on the $\nu=2$ plateau are plotted vs. $B$. On average, $\Rds$ varies around zero, indicating that the CAES are roughly equal superpositions of electrons and holes. 
\end{figure}

\section*{Dependence of $\Rds$ on magnetic field}
The behavior observed in Fig.\,1C is generic. In Fig.\,2A, we plot $\Rds$ and $G_{xy}$ at $3$ T in a wider range of gate voltages. Clearly, $\Rds$ oscillates around zero as a function of the gate voltage for a range of integer filling factors. We could also expect mesoscopic fluctuations in the downstream resistance to be induced by changing the magnetic field, $B$, which changes the magnetic length. Indeed, our simulations indicate that $P_e-P_h$ oscillates with magnetic field (Fig.\,S13).

To explore this dependence experimentally, we plot $\Rds$ as a function of both $B$ and $V_G$ in Fig.\,2B. The overall pattern is reminiscent of the traditional Landau fan diagram of the longitudinal resistance, with the exception that the downstream resistance $\Rds$ is not equal to zero on the plateaus. Mesoscopic fluctuations that deviate from zero resistance (white) appear inside the quantum Hall plateaus as blue (negative) and red (positive) pockets. A prominent feature of this data is the frequent abrupt changes of the $\Rds(V_G)$ pattern while sweeping the magnetic field.  Although the field sweeps are stable and reproducible in a very small range of $B$, changing the field by several mT completely changes the $\Rds(V_G)$ curves (see Fig.\,S5B for more details). This stochastic switching complicates the analysis of the map.

We note that when switching of the $\Rds(V_G)$ curve occurs, the $G_{xy}$ stays unchanged (compare Fig.\,1B and 2B). This can also be noticed in Fig.\,S5A in the regions slightly outside the quantized plateau, where $G_{xy}$ develops a recognizable mesoscopic pattern. This observation indicates that the switching events do not involve the normal contacts \textbf{\textit{b}} and \textbf{\textit{d}} between which $G_{xy}$ is measured, nor the bulk of the sample. Instead, they must originate in the superconducting contact \textbf{\textit{c}}. We surmise that the switching of the $\Rds$ pattern is caused by the rearrangement of vortices inside the type II superconducting contact. Indeed, we have routinely observed similar switching events in the interference pattern of supercurrent in Josephson junctions fabricated with similar contacts. In Supplementary Information section S4, we show that we can hysteretically switch between two distinct patterns of $\Rds$ multiple times, indicating that the vortices can be controllably added to and removed from the superconductor.

To explain the observed sensitivity of $\Rds$ to the vortex configuration, we note that adding a vortex close to the interface should change the phase of the order parameter along the quantum Hall-superconductor interface $\theta (x)$. As a result of this change, a pure electron or a pure hole state would only acquire an overall phase, which would not change $\Rds$. However, the change of $\theta (x)$ is expected to change the relative phase shift between the two interfering CAES. 
The presence of the vortices is typically neglected in theoretical studies, but we find that they have a dramatic effect on the beating pattern of the CAES -- in Fig.\,S12 we show that the $P_e-P_h$ curve is completely scrambled by adding just a single vortex, modeled as a kink in $\theta (x)$.

To extract information otherwise buried in the stochastic switching, we analyze the impact of $B$ on the gate-dependent oscillations on the $\nu=2$ plateau (more statistical analysis can be found in Supplementary Information section S7). The plateau region used for this analysis is selected such that $G_{xy}$ is within 1\% off the quantized values, as indicated by the black lines in Fig.\,2B. We then find the minimum and maximum $\Rds(V_G)$ for a given field and plot the resulting $\min\Rds(V_G)$ and $\max \Rds(V_G)$ as a function of $B$ in Fig.\,2C. First, we find that the amplitude of the fluctuations decreases with $B$. Most likely, this suppression is explained by the CAES being absorbed by the contact, thereby creating quasiparticle excitations in the superconductor. (These excitations are possibly absorbed by the normal cores of the vortices.) Evidently, this process becomes more effective at higher $B$. 
Second, the typical amplitudes of the positive and negative signals are very close. We argue that this observation indicates that CAES are on average neutral (see Supplementary Information section S2\&S6). Indeed, if the two CAES $\psi_{1,2}$ had predominantly electron-like and hole-like characters, the incoming electron would couple primarily to $\psi_1$. This would in turn result in a greater likelihood of electrons being emitted downstream, and $\Rds$ would mostly stay positive, i.e. $\max\Rds(V_G) > |\min \Rds(V_G)|$, contrary to our observations. 

Our numerical simulations support this argument: the eigenmodes $\psi_{1,2}$ are given by coherent superpositions of electron and hole amplitudes, which have distinct patterns in space. Nevertheless, the integral of the probability of the electron and hole components is close to $1/2$ (Fig.\,S11), resulting in the overall approximately neutral character of the CAES. Due to particle-hole symmetry of the model, $\psi_2$ at zero energy is the charge conjugate partner of $\psi_1$, meaning that the pattern of the electron and hole amplitudes is interchanged. 

\begin{figure}
\includegraphics[width=1\textwidth]{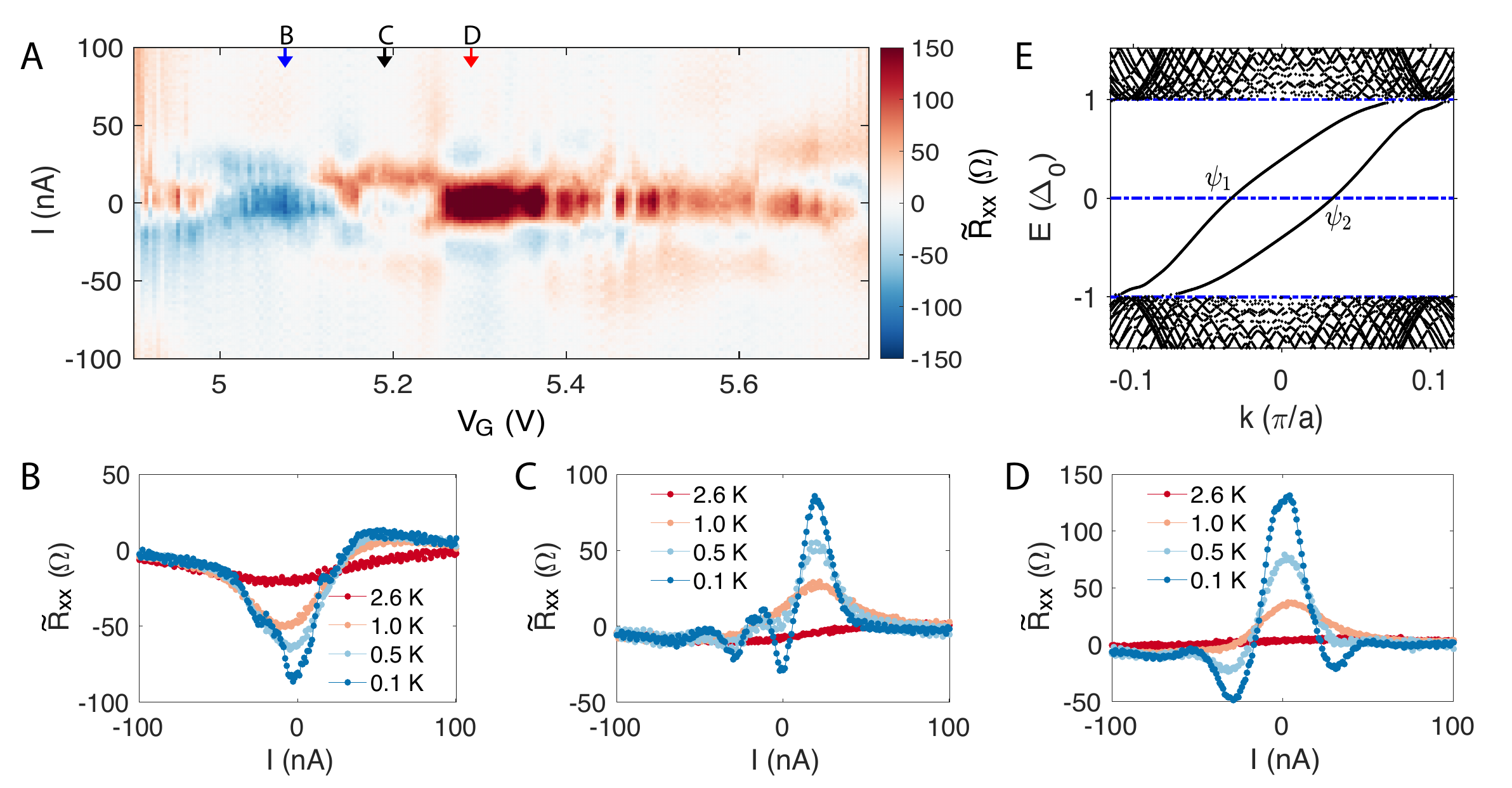}

\noindent {\bf Fig.\,3.} {\bf The bias-dependence of the interference effect.} (\textbf{A}) The superconductor downstream longitudinal resistance $\Rds$ plotted versus the DC bias current $I$ and gate voltage $V_G$ on a well-quantized $\nu=2$ plateau at $B=$ 3 T. Gate-dependent oscillations centered at zero-bias are observed, indicating interference of the CAES. (\textbf{B-D}) Bias-dependent oscillations of $\Rds$ at the gate voltages $V_G=5.08$ (\textbf{B}), $5.19$ (\textbf{C}) and $5.29$ (\textbf{D}) V marked by the arrows in (\textbf{A}). Oscillations die out with increasing temperature. (\textbf{E}) The dispersion relation of a pair of CAES at a quantum Hall-superconductor interface calculated from the tight-binding simulation. The momentum difference between the two modes varies with energy, causing oscillations of $\Rds$ in bias.
\end{figure}

\section*{Dispersion of CAES}
Finally, we address the non-linearity of the CAES energy-momentum dispersion. In Fig.\,3A, we plot the $\Rds$ map measured as a function of $I$ and $V_{G}$. This data corresponds to the range nominally identical to Fig.\,1C; however it was measured following a sweep of magnetic field, so the individual mesoscopic features have changed. The dependence of this map on temperature and additional maps for $\nu=6$ are shown in section S5 of Supplementary Information.

Most notably, we find that $\Rds$ in Fig.\,3A oscillates not only with the gate voltage, but also as a function of the DC current bias. These oscillations are revealed in the vertical cross-sections of the map, taken at $V_G=5.08, 5.19$ and $5.29$ V and plotted in Fig.\,3B-D. Quite unusually, the signal can even oscillate several times as a function of bias, as shown in Fig.\,3C. To interpret these oscillations, we note that the applied current tunes the energy of the injected electrons with respect to the grounded superconducting contact, $E=eI/G_{xy}$. The wavevector difference $\delta k$ between the two CAES depends on their energy, as demonstrated in our model calculations in Fig.\,3E (see also Supplementary Information section S6). The phase difference accumulated by the CAES along the interface, $\delta k L$, thus produces the observed bias oscillations of $\Rds$. Eventually, $\Rds$ goes to zero when the applied voltage becomes comparable to the superconducting gap, $I/G_{xy} \approx \Delta/e$. At that point, the incoming electrons have high probability to enter the superconductor as quasi-particles, and no downstream signal is expected. 

\section*{Outlook}
We have demonstrated robust coupling of the quantum Hall edge states to a superconductor via Andreev reflections, resulting in the formation of chiral Andreev edge states - coherent superpositions of electrons and holes. Further study of the CAES may focus on increasing the strength of the downstream signal, and making the oscillation pattern more regular, both of which could be achieved by shortening the superconducting interface and reducing magnetic field. Such developments could in turn lead to realization of novel quantum effects and devices, the possible examples being a Bogoliubov quasi-particle annihilation~\cite{beenakker_annihilation_2014}, a superconducting flux capacitor~\cite{clarke_exotic_2014}, and a phase-coherent heat circulator~\cite{heatcirculator}. Finally, the non-local downstream voltage measurement implemented here could be further applied to samples in which the quantum Hall is replaced by the quantum anomalous Hall. In this case, the neutral interfacial modes are predicted to be chiral Majorana fermions~\cite{lian_edge-state-induced_2016}.

\section*{Acknowledgments}
We greatly appreciate stimulating discussion with Albert Chang, Matthew Gilbert, Biao Lian, Yuval Oreg, Kirill Shtengel, and Ady Stern. {\bf Funding:} 
Transport measurements conducted by L.Z., E.G.A., and A.S. were supported by Division of Materials Sciences and Engineering, Office of Basic Energy Sciences, U.S. Department of Energy, under Award No. DE-SC0002765. 
Lithographic fabrication and characterization of the samples was performed by L.Z. and A.S. with the support of NSF awards ECCS-1610213 and DMR-1743907. 
The measurement setup was developed by A.W.D., T.L., and G.F. with the support of ARO Award W911NF-16-1-0122. 
Numerical simulations conducted by A.B. and H.U.B. were supported by Division of Materials Sciences and Engineering, Office of Basic Energy Sciences, U.S. Department of Energy, under Award No. DE-SC0005237. 
H.L. and F.A. acknowledge the ARO under Award W911NF-16-1-0132.
K.W. and T.T. acknowledge support from JSPS KAKENHI Grant Number JP15K21722 and the Elemental Strategy Initiative conducted by the MEXT, Japan. 
T.T. acknowledges support from JSPS Grant-in-Aid for Scientific Research A (No. 26248061) and JSPS Innovative Areas Nano Informatics (No. 25106006).
The sample fabrication was performed in part at the Duke University Shared Materials Instrumentation Facility (SMIF), a member of the North Carolina Research Triangle Nanotechnology Network (RTNN), which is supported by the National Science Foundation (Grant ECCS-1542015) as part of the National Nanotechnology Coordinated Infrastructure (NNCI).

\section*{Author contributions}
L.Z. and A.S. characterized and fabricated the device. H.L. and F.A. made the graphene-hBN heterostructrue. T.T. and K.W. provided the hBN crystals. L.Z., E.G.A. and A.S. performed the measurements. A.W.D., T.L. and G.F. developed the measurement setup. A.B. and H.U.B. conducted the numerical calculations. L.Z. and G.F. analyzed the data and wrote the manuscript. H.U.B., F.A. and G.F. supervised the project. 

\section*{Competing Interests}
The authors declare no competing interests.

\section*{Methods}

\subsection*{Device Fabrication}

The heterostructure is assembled from separately exfoliated flakes of graphene and hexagonal boron nitride (hBN). These flakes are deposited onto a diced silicon wafer capped with a 280 nm thermally grown oxide layer. The substrate has been baked to remove moisture (200 $^\circ$C for 10 minutes) and oxygen plasma ashed (10 s at 500 mbar) in the case of hBN. 

The flakes are assembled with the dry transfer technique: they are picked up with a polydimethylsiloxane (PDMS)/polycarbonate (PC) stamp. First, a 2x2 mm square of PDMS is cut and attached to a glass slide with transparent, double-sided tape. Meanwhile, PC film (6 g suspended per 100 ml chloroform) is prepared by drop-coating a separate slide and leveling the layer by dragging another clean slide across it, before leaving the solvent to evaporate. The PC layer then is picked up with double-sided tape and placed (tape-side down) over the PDMS square. The resulting PDMS/PC stamp is then cured for 5 minutes at 110 $^\circ$C.
 
Individual flakes are picked up with the stamp at 70 $^\circ$C and then deposited onto a clean Si/SiO2 substrate at 150 $^\circ$C. The finished stack is cleaned of stamp residues in hot dichloromethane for 10 minutes followed by a 30 minute anneal step at 500 $^\circ$C in atmosphere. This has the additional benefit of consolidating the ``bubbles'' of trapped hydrocarbons within the heterostructure, leaving larger defect-free regions. The clean region of encapsulated graphene used for the device is identified using atomic force microscopy and Raman spectroscopy mapping. 

The patterning of the heterostructure is achieved by electron-beam lithography on a layer of PMMA resist. We use SF$_6$ to etch the top layer of hBN followed by CHF$_3$/O$_2$ to etch the graphene layer. The electrodes are deposited right after etching the contact region, to ensure a fresh graphene-metal interface. The normal metal electrodes are made of thermally evaporated Cr (1 nm)/Au (100 nm). The superconducting electrodes are 100 nm molybdenum rhenium alloy (50-50 ratio by weight) DC sputtered in a high vacuum chamber ($10^{-8}$ Torr).    

\subsection*{Measurements}
Measurements were performed in a Leiden Cryogenics dilution refrigerator at a temperature of 100 mK unless otherwise stated. The sample was connected in the refrigerator via resistive coax lines and low-temperature RC filters. Differential resistance measurements were carried out using a 10 nA square wave excitation with home-made preamplifiers and a NI-6363 data acquisition board. The excitation frequency was chosen to be 15 Hz in order to maximize 60 Hz noise reduction. The response to the square wave step was allowed to settle for a time greater than the RC time of the filters before recording the voltage. 

\subsection*{Data Availability}
Source data for figures (including Supplementary Figures) are available in the public repository Zenodo, http://doi.org/10.5281/zenodo.3708374~\cite{CAESdatacodes}. All other data that support the plots within this paper and other findings of this study are available from the corresponding author upon reasonable request.

\subsection*{Code Availability}
The codes used for the analysis and simulations are available in the public repository Zenodo, http://doi.org/10.5281/zenodo.3708374~\cite{CAESdatacodes}.

\section*{Supplementary Information}

\global\long\def\theequation{S\arabic{equation}}
\global\long\def\thefigure{S\arabic{figure}}
\global\long\def\thesection{S\arabic{section}}
\setcounter{equation}{0}
\setcounter{figure}{0}
\setcounter{section}{0}

\subsection*{S1. Deviations of $\Rds$ from the conventional longitudinal resistance}

\begin{figure}
\centering
\includegraphics[width=1\textwidth]{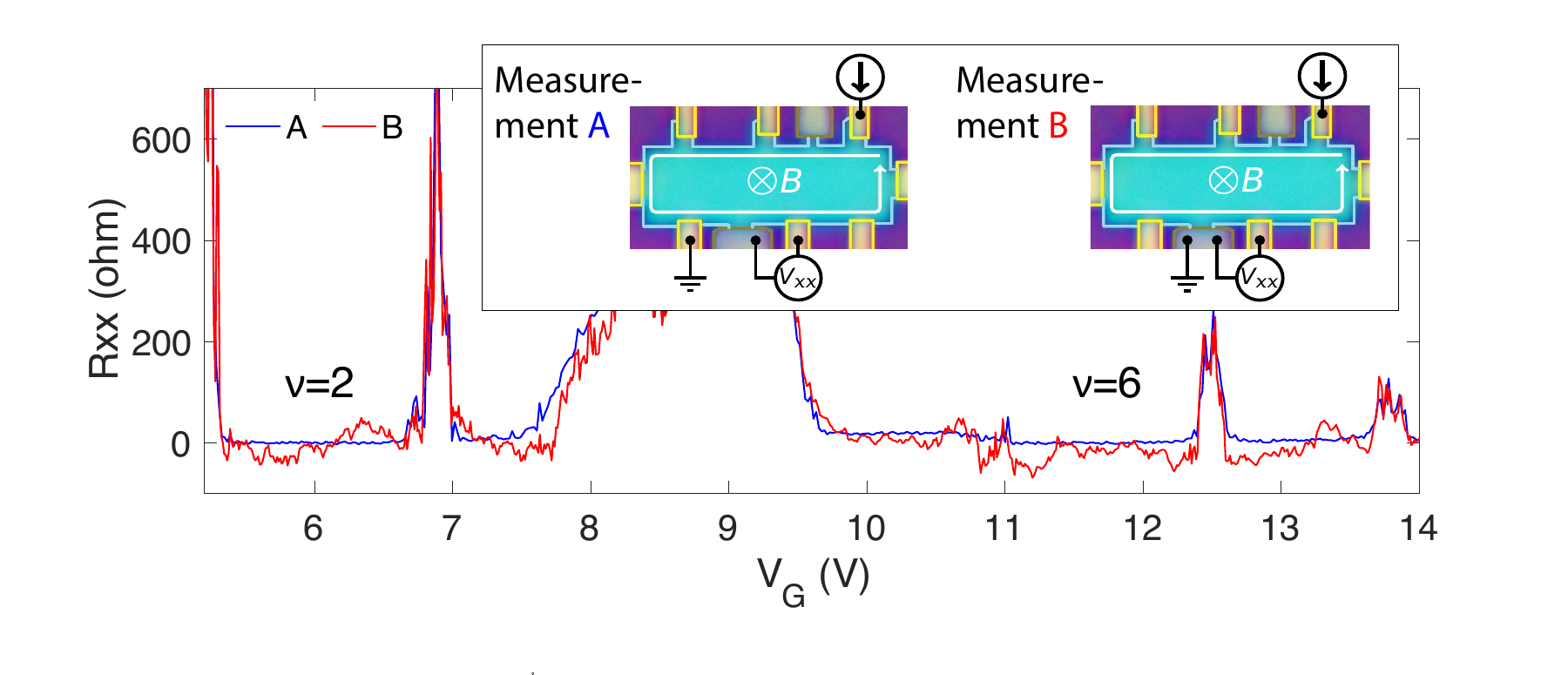}
\caption{
Longitudinal resistance measured at $B=4$ T as a function of gate voltage $V_G$ in two configurations: A (blue) and B (red). Configuration B is identical to the one used elsewhere in the text; configuration A is conventional for quantum Hall measurements: the voltage is measured between two floating contacts. Note that in both cases the source and the voltage probes are the same, but in A the drain is moved further to the left (upstream) from the superconducting contact that was grounded in B. Configuration A results in vanishing $R_{xx}$, as expected.}
\label{figs1}
\end{figure}

\begin{figure}
\centering
\includegraphics[width=1\textwidth]{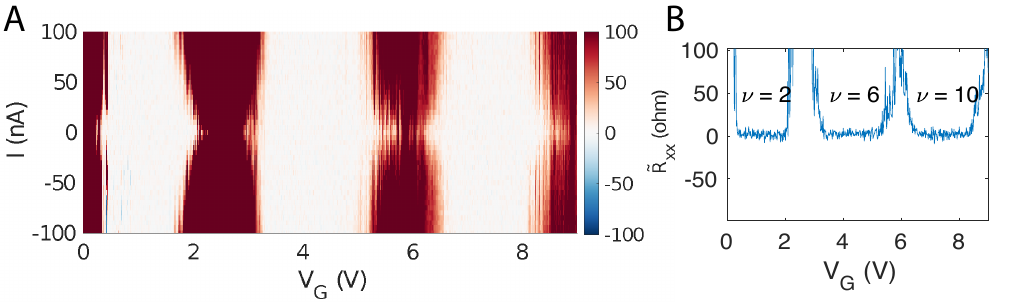}
\caption{
$\Rds$ measured downstream from a normal contact at 2 Tesla. The white regions (panel A) of zero $\Rds$ correspond to filling factors 2, 6 and 10. The zero-bias cut is presented in panel B.
}
\label{figNDS}
\end{figure}

To rule out a bulk contribution to the non-zero $\Rds$ on the quantum Hall (QH) plateaus, we have measured both the $\Rds$ and the conventional longitudinal resistance $R_{xx}$ using very similar setups. The inset of Fig.\,\ref{figs1} shows the measurement configurations for $R_{xx}$ (A) and for $\Rds$ (B). In comparison with scheme B, the drain in scheme A is moved from the superconductor (SC) to the normal metal lead located further upstream. In this way, the bulk contribution to $V_{xx}$ is almost the same because the voltage probes and the source remain unchanged, and the position of the drain moves only slightly. Fig.\,\ref{figs1} shows that $R_{xx}$ and $\Rds$ nearly coincide away from the quantum Hall plateaus, where the contribution of the bulk states is most pronounced. They significantly differ in the plateau regions where $R_{xx}$ is nearly zero. We conclude that the non-zero $\Rds$ on the plateaus is the result of the influence of superconductor on the quantum Hall edge states which we recognize as the CAES.  

To further stress the point that the non-trivial $\Rds$ originates at the grounded superconducting interface, we measure a similar sample with normal contacts. Fig.\,\ref{figNDS} shows $\Rds$ measured for the normal contact with the interface length of 1 $\mu$m, comparable to the graphene-superconductor interface length studied in the paper. No discernable deviations from zero resistance are observed in this case.

\begin{figure}
\centering
\includegraphics[width=1\textwidth]{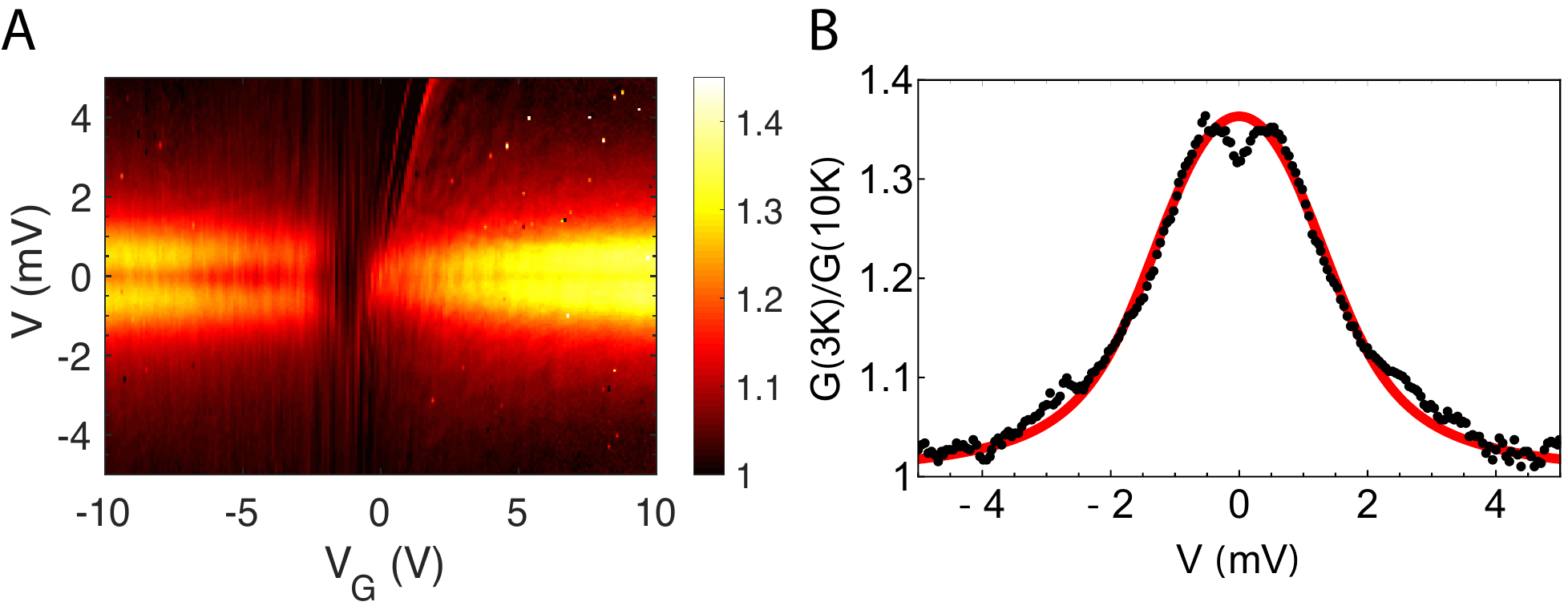}
\caption{
Differential conductance of the graphene-MoRe interface measured at $B=0$ and $T=3$ K, normalized by the high temperature conductance, $G(V,3K)/G(V,10K)$. (A) Normalized conductance plotted versus bias voltage $V$ and gate voltage $V_G$. Low-bias conductance enhancement is observed at low temperatures through most of the gate voltage range. (B) Normalized conductance versus bias at $V_G=10$ V. The red curve is calculated using the modified Blonder-Tinkham-Klapwijk theory~\cite{S_lee_inducing_2017} by taking the superconducting gap $\Delta=1.3$ meV, energy broadening $\Gamma=0.5\Delta$, barrier parameter $Z=0$ and temperature $T=3$ K. (Here, we use a different sample than the one presented in the main text; however, the contacts are fabricated following the same procedure.) 
}
\label{figBTK}
\end{figure}

To rule out effects from non-transparent interfaces, we have studied a similar device (graphene with a MoRe contact and normal contacts). The differential conductance of the interface is measured in a properly filtered 3 K cryostat and normalized by the 10 K (above the $T_c \approx 9$ K) data. Fig. \ref{figBTK} plots the normalized differential conductance at zero magnetic field as a function of the bias voltage and gate voltage. Enhanced conductance inside the superconducting gap is universally observed away from the Dirac point ($V_G=-1.3$ V). In the well n-doped region ($V_G>5$ V), the enhancement factor is above 1.3 suggesting a highly transparent contact (note that the enhancement should be even higher at lower temperatures). We fit the normalized differential conductance at $V_G=10$ V to the modified Blonder-Tinkham-Klapwijk model \cite{S_lee_inducing_2017} (see panel B) and find the superconducting gap $\Delta=1.3$ meV, energy broadening $\Gamma=0.5\Delta$ and barrier parameter $Z=0$. These findings also agree with a transparent graphene-MoRe interface. (A small barrier is likely present, evidenced by a small dip at zero bias, but the fitting cannot reliably determine its value.)

\subsection*{S2. Toy model of the interference of chiral Andreev edge states} 

For a spinless edge state, the s-wave superconducting proximity effect can be described by the Bogoliubov-de Gennes Hamiltonian in the basis of $\{\ket{e},\ket{h}\}$~\cite{S_van_ostaay_spin-triplet_2011}. Its particle-hole symmetry ensures that the two eigenstates at zero energy can be written as~\cite{S_lian_edge-state-induced_2016}
 \begin{equation}
 \begin{aligned}
 &\ket{\psi_1}= \alpha \ket{e} + \beta \ket{h} \\
 &\ket{\psi_2}= \beta^* \ket{e} - \alpha^* \ket{h} ,
 \end{aligned}
 \end{equation}
where $|\alpha|^2+|\beta|^2=1$. An incoming electron $\ket{e}=\alpha^* \ket{\psi_1}+\beta\ket{\psi_2}$ propagates through the proximity region as 
\begin{equation}
\begin{aligned}
\ket{\phi}&=\alpha^* e^{i k_1 L} \ket{\psi_1}+\beta e^{i k_2 L} \ket{\psi_2}\\
&=\left( |\alpha|^2 e^{i k_1 L}+|\beta|^2 e^{i k_2 L}\right) \ket{e} 
+(e^{i k_1 L}-e^{i k_2 L}) \alpha^* \beta \ket{h},
\end{aligned}
\end{equation}
where $k_1$, $k_2$ are the wavevectors of the two modes and $L$ is the propagation length. The probability of converting an electron into a hole is then $P_h = 4 |\alpha|^2|\beta|^2 \sin^2(\delta k L/2)$, where 
$\delta k=k_1-k_2$, so that $\delta k L$ is the acquired phase difference between the two modes. Thus the charge coming out of the CAES is
\begin{equation}
q=e(P_e-P_h)=e(1-2 P_h)=e\left(1-8|\alpha|^2|\beta|^2 \sin^2(\delta k L/2)\right).
\end{equation}
Averaging over the phase difference, we find 
\begin{equation}
\overline{q}=e (1-4|\alpha|^2|\beta|^2).
\end{equation}
Notice that the average result is neutral when the two eigenstates are neutral ($|\alpha|^2\!=\!|\beta|^2\!=\!1/2$): when the interface modes are neutral electron-hole hybrids, an incoming electron is equally likely to produce an outgoing hole as an outgoing electron. 

In the $\nu\!=\!2$ quantum Hall state that is the focus of this paper, the spins of the electrons are not polarized, and so there are two possible spin states for the electrons and two for the holes. Thus, there are four eigenstates of the Bogoliubov-de Gennes Hamiltonian at zero energy, which are pairwise charge conjugate to each other. For the argument above, we have assumed that both $\ket{\psi_1}$ and $\ket{\psi_2}$ are spin degenerate. In a more sophisticated analysis this does not have to be the case.

\subsection*{S3. Landauer-B\"uttiker picture}

\begin{figure}
\centering
\includegraphics[width=0.5\textwidth]{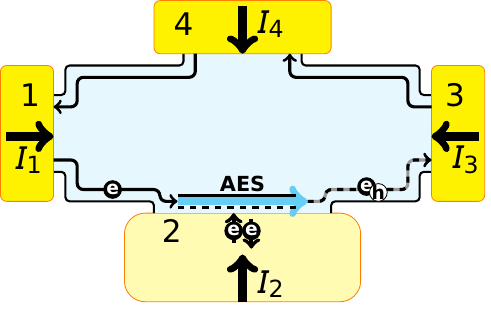}
\caption{
Sketch of the four-terminal transport measurement, in which the bottom contact 2 is superconducting and the others are normal. An electron injected into the CAES (bottom left) comes out as a hole or electron (bottom right) with a probability $P_h$ or $P_e$.}
\label{figLB}
\end{figure}

As shown in Fig.\,\ref{figLB}, a two-dimensional electron gas device with $\nu$ chiral channels is contacted with four metal leads. The bottom lead is a grounded superconductor and the other three are normal metal. We calculate the zero-bias Hall conductance $G_{xy}$ and superconductor downstream resistance $\Rds$ at zero temperature using the Landauer-B\"uttiker formula \cite{Ihnbook} assuming 
(i) no electron can transfer into the superconductor without forming a Cooper pair and (ii) an electron injected into the CAES comes out as a hole (or electron) with a probability $P_h$ (or $P_e$). The relation between the current in the leads $I_i$ and the voltage on them $V_j$ is
\begin{equation}
\left( \begin{array}{c} I_1\\I_2\\I_3\\I_4 \end{array} \right)=
\nu \frac{e^2}{h}
\left( 
\begin{array}{cccc}
1 & 0 & 0 & -1\\
-1+(P_e-P_h) & 1-(P_e-P_h) & 0 & 0\\
-(P_e-P_h) & -1+(P_e-P_h) & 1 & 0\\
0 & 0 & -1 & 1
\end{array}
\right)
\left( \begin{array}{c} V_1\\V_2\\V_3\\V_4 \end{array} \right)=
\left( \begin{array}{c} 0\\-I\\0\\I \end{array} \right).
\label{eqGmaxtrix}
\end{equation}

Deriving the conductance matrix in the superconducting case requires extra care due to the Andreev processes. Here we derive the individual lines in Eq.\,\ref{eqGmaxtrix} following reference \cite{S_takagaki_transport_1998}. The net currents flowing through contacts 1 and 4 are unaffected by superconductivity. The currents affected by Andreev reflections are $I_2$ and $I_3$. Let us refer all the voltages with respect to the grounded superconducting contact 2. $I_2$ is the current of Cooper pairs flowing out of the contact 2. It can be calculated as a current of electrons coming from contact 1, Andreev reflected with a probability $P_h$:
\begin{equation}
I_2=-\frac{\nu e^2}{h}(V_1-V_2)2P_h
\end{equation}
The factor of 2 indicates that for each Andreev reflection of an electron to a hole, a Cooper pair is added to the superconducting contact. Rewriting $2P_h$ in the symmetric form $1-P_e+P_h$ gives the second line in Eq.\,\ref{eqGmaxtrix}.

$I_3$ consists of the trivial current flowing downstream from contact 3, minus the non-trivial current flowing to contact 3 from the upstream direction. The latter term is calculated from the probability of an electron from contact 1 to flow past contact 2 as an electron or to turn into a hole: $\frac{\nu e^2}{h}(P_e-P_h)(V_1-V_2)$. (Alternatively, one can think of this term as the difference of currents flowing from contact 1 and the current absorbed by contact 2.) Overall,
\begin{equation}
I_3=\frac{\nu e^2}{h}\left((V_3-V_2)-(P_e-P_h)(V_1-V_2)\right)
\end{equation}

We can now check the consistency of the resulting matrix. Current conservation dictates that the sum of the elements in a column must be equal to zero. The sum of the elements in every row should also be zero: if we apply a uniform change of the chemical potential to all contacts, the currents through any contact should remain the same. The conductance matrix indeed satisfies both sets of conditions.

Solving Eq.\,\ref{eqGmaxtrix}, we find the Hall conductance $G_{xy}$ and superconductor downstream resistance $\Rds$
\begin{equation}
\begin{aligned}
G_{xy}&=\frac{I}{V_1-V_3}=\nu \frac{e^2}{h}\\[0.2cm]
\Rds&=\frac{V_3-V_2}{I}=\frac{P_e-P_h}{1-(P_e-P_h)}G_{xy}^{-1}.
\end{aligned}
\label{eq:RDS}
\end{equation}

The CAES may be absorbed by the superconductor as quasi-particle excitations, reducing both $P_e$ and $P_h$. We introduce a phenomenological ``survival probability'' $P_{\mathrm{surv.}}$ to describe the fraction of the particles reaching the end of the interface. We then multiply $P_e-P_h$ by $P_{\mathrm{surv.}}$, and the superconductor downstream longitudinal resistance $\Rds$ becomes

\begin{equation}
\Rds=\frac{V_3-V_2}{I}=\frac{(P_e-P_h)P_\mathrm{surv.}}{1-(P_e-P_h)P_\mathrm{surv.}}G_{xy}^{-1}.
\end{equation} 

Note that since in the experiment $\Rds \ll h/e^2$, the survival probability has to be smaller than 1. Under this condition, $\Rds\approx (P_e-P_h)P_\mathrm{surv.}G_{xy}^{-1}.$

\subsection*{S4. Stochastic switching behavior}

\begin{figure}[t]
\centering
\includegraphics[width=1\textwidth]{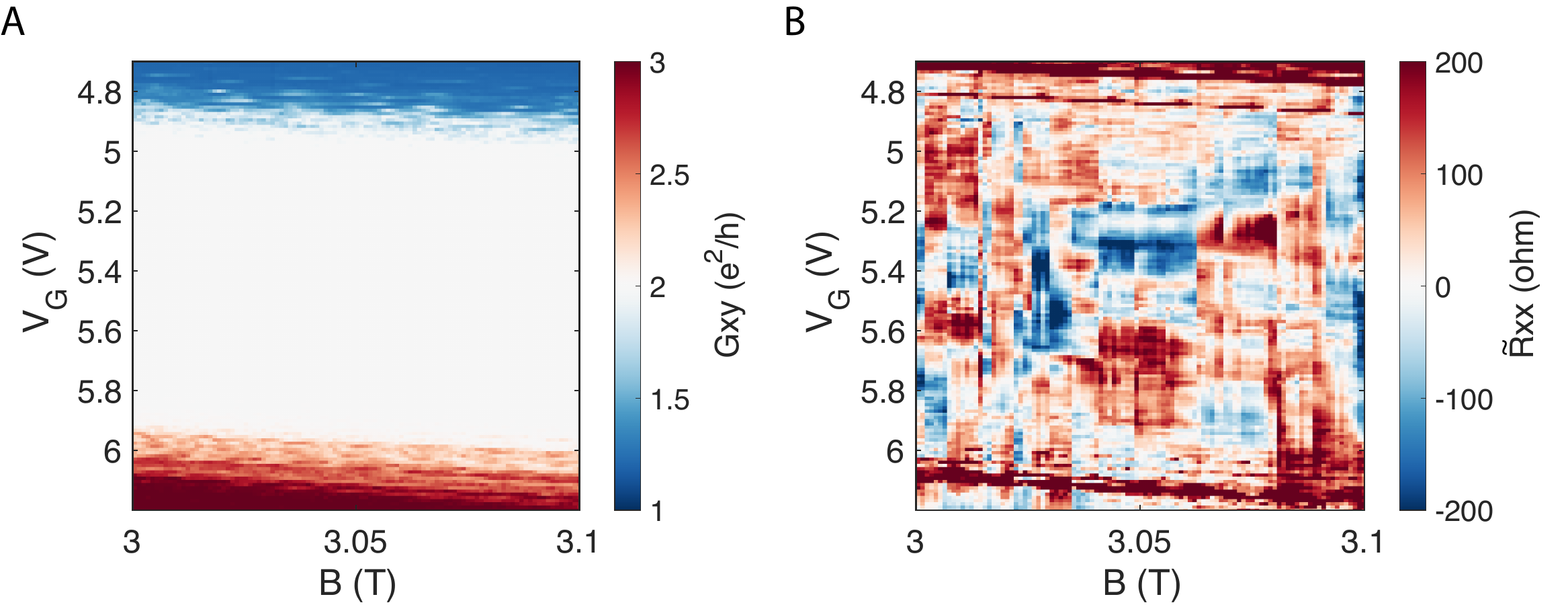}
\caption{
Detailed maps of zero-bias $G_{xy}$ (\textbf{A}) and $\Rds$ (\textbf{B}) simultaneously measured while ramping the field down from 3.1 T to 3 T.}
\label{figS3}
\end{figure}

As discussed in the main text, the downstream resistance $\Rds$ measured on top of well quantized $G_{xy}$ plateaus shows stochastic switching behavior. Fig.\,\ref{figS3} presents the details of this switching by zooming into a small region of Fig.\,2B. Zero-bias $G_{xy}(V_G)$ and $\Rds(V_G)$ are plotted in a field range of 3 to 3.1 T on top of the $\nu=2$ plateau. Clearly, $\Rds$ exhibits switching behavior (vertical lines) while $G_{xy}$ remains smooth, as seen by the continuity of the inclined mesoscopic features at the transition between the plateaus. $G_{xy}$ uses the same normal contacts (\textbf{\textit{a}} and \textbf{\textit{d}} in Fig.\,1A) and probes the same region of the sample as $\Rds$; the only difference is that the measurement of $\Rds$ involves the voltage on the superconducting contact \textbf{\textit{c}}. We thus attribute the switching to the rearrangement of vortices in that contact.

\begin{figure}
\centering
\includegraphics[width=0.7\textwidth]{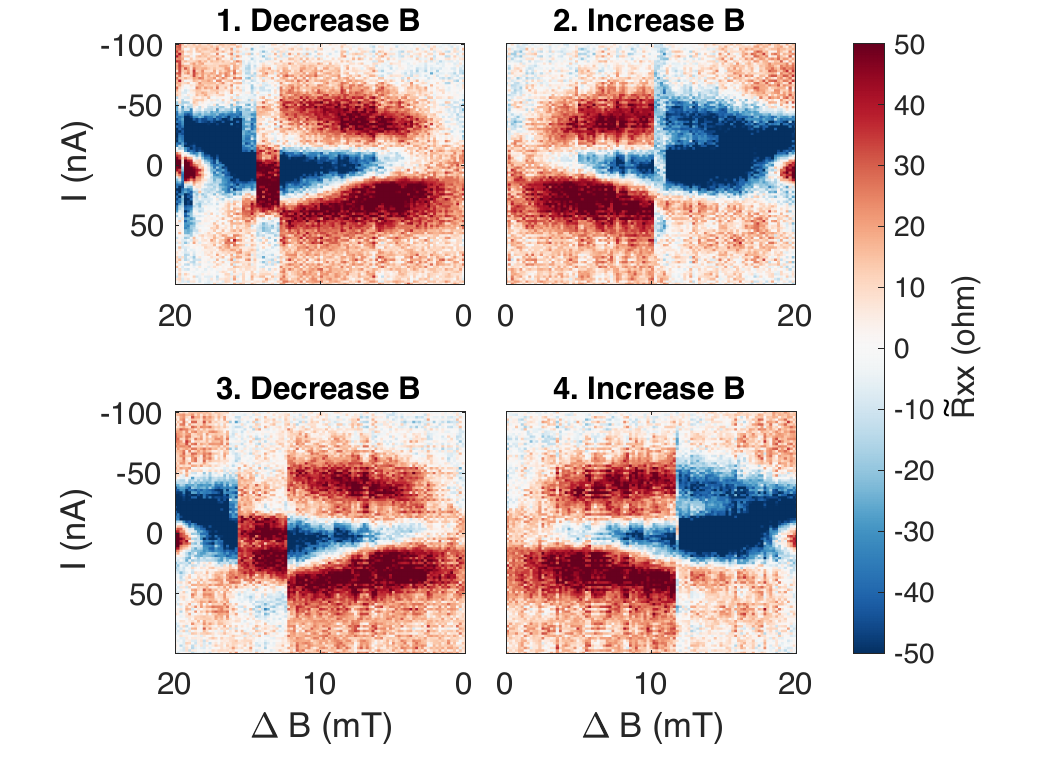}
\caption{
The downstream resistance $\Rds$ measured at $V_G=5.67$ V as a function of the bias current and $\Delta B$, the deviation of the magnetic field from 3 T. The 4 panels correspond to successive field sweeps from 3.02 T to 3 T and up again, repeated twice. }
\label{figvortex}
\end{figure}

To support this argument, we measure $\Rds$ as a function of bias at a fixed gate voltage $V_G=5.67$ V while sweeping the magnetic field back and forth near 3 T. As shown in the first quadrant of Fig.\,\ref{figvortex}, we start by sweeping the magnetic field from 3.02 T to 3 T. During this process, $\Rds$ switches from a dip at zero bias to a peak and then back to a dip around 3.013 T. Then we sweep the magnetic field from 3.02 T back to 3 T (the second quadrant), during which $\Rds$ directly switches from a zero-bias dip to another dip pattern around 3.01 T. At the end of the cycle, $\Rds$ goes back to the initial state at 3.02 T. The remaining two plots show successive sweeps down and up, which produce very similar patterns.

The arrangement of vortices in the superconductor near the interface can strongly influence the phases of the interface modes. This is particularly clear in the semiclassical picture where a particle skipping along the edge picks up a phase from the superconductor upon each Andreev reflection. These phases are determined by the arrangement of vortices. In the quantum picture, such as $\nu\!=\!2$ studied here, the beating between the CAES changes due to these phases. Therefore, the data presented in Fig.\,\ref{figvortex} can be interpreted as follows. 
As the magnetic field decreases from about $3.015$ T to $3.01$ T, two vortices appear to be removed one by one, generating a dip-peak-dip switching feature in $\Rds$. The two vortices are apparently added back at the same time when we increase the magnetic field (panel 2) so that only one switch is observed at $B\approx 3.01$ T. 

\subsection*{S5. More data of the interference of CAES}

\begin{figure}[hb]
\centering
\includegraphics[width=0.85\textwidth]{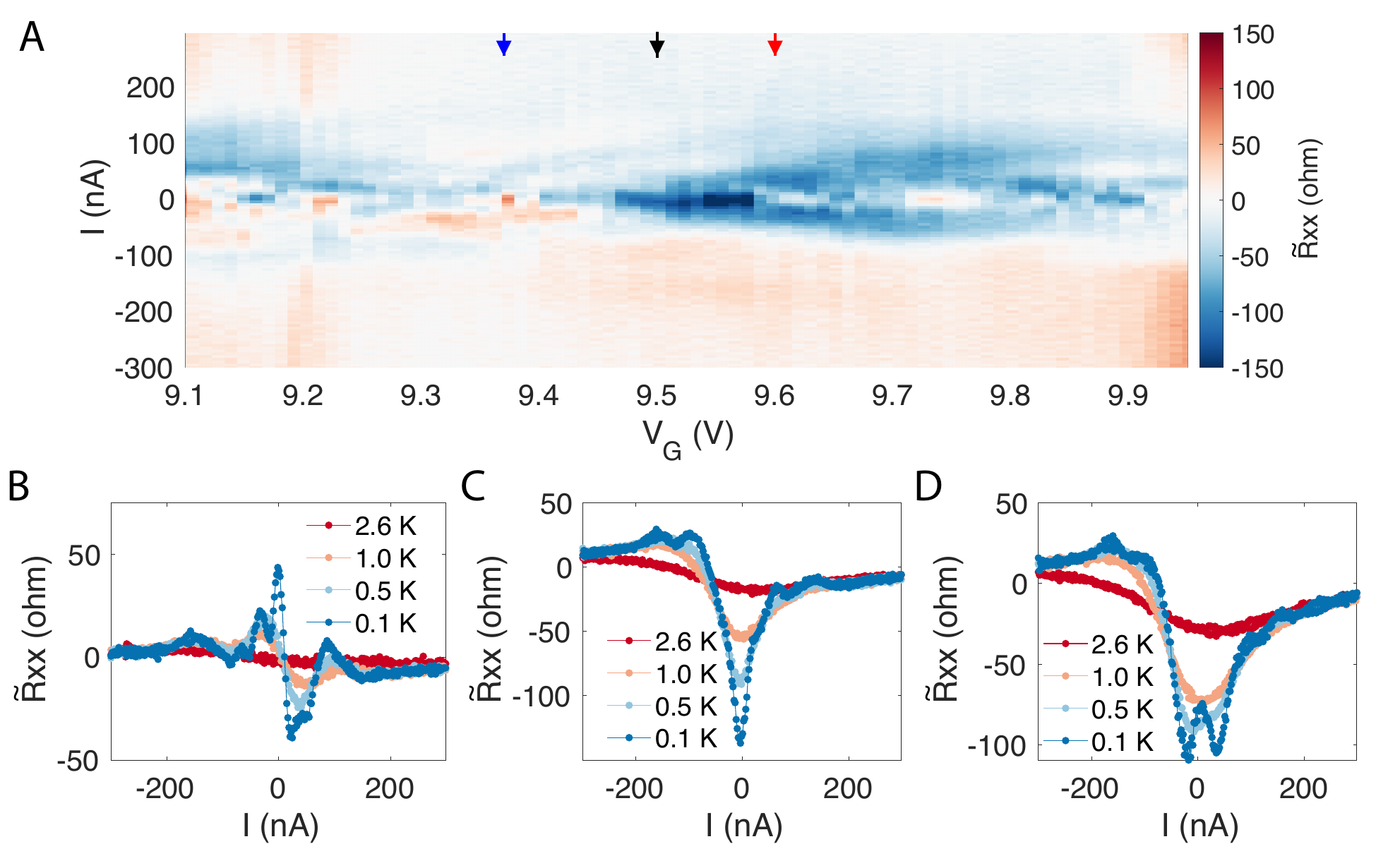}
\caption{
(\textbf{A}) The superconductor downstream longitudinal resistance $\Rds$ plotted versus the DC bias current $I_\text{bias}$ and gate voltage $V_G$ on the $\nu=6$ plateau at $B=$ 3 T. Gate-dependent oscillations centered at zero-bias are observed inside the well-quantized region, indicating interference of the CAES. (\textbf{B-D}) Bias-dependent oscillations of $\Rds$ at the gate voltages $V_G=9.37$ (\textbf{B}), $9.50$ (\textbf{C}), $9.60$ (\textbf{D}) V marked by the arrows in (\textbf{A}).}
\label{figv6int}
\end{figure}

\begin{figure}
\centering
\includegraphics[width=0.45\textwidth]{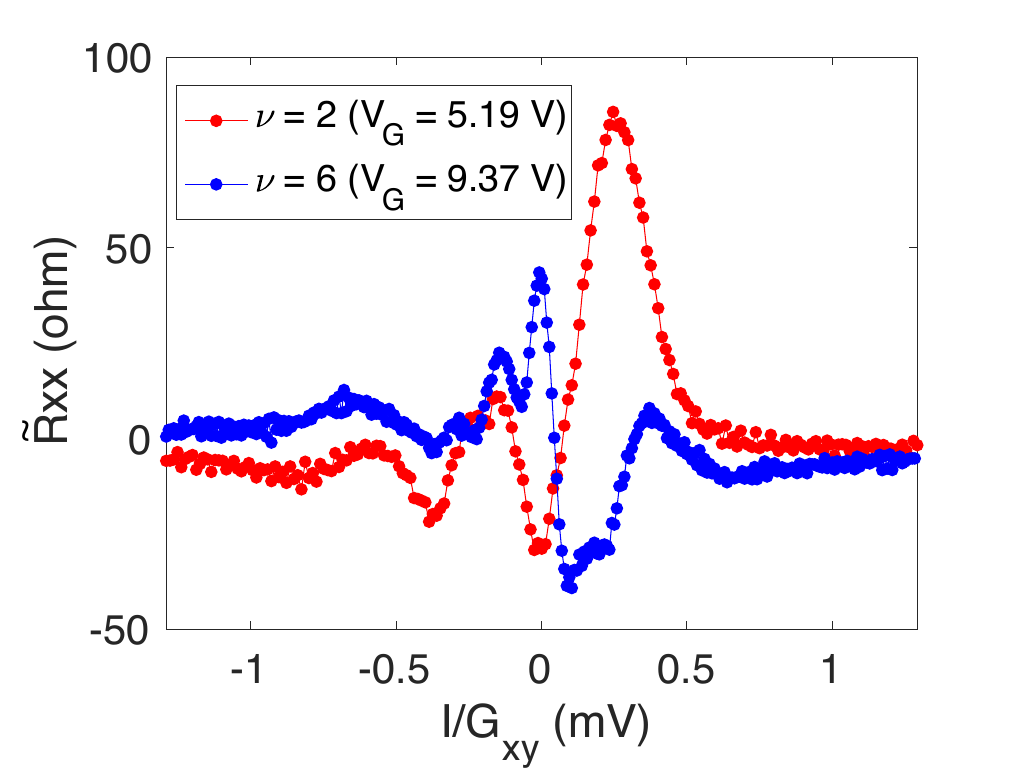}
\caption{
Comparison of $\Rds$ plotted vs. bias voltage at two filling factors $\nu=2$ and $6$. }
\label{figcutv26}
\end{figure}

To complement Fig.\,3, in Fig.\,\ref{figv6int} we show the bias-gate oscillations of $\Rds$ on the $\nu=6$ plateau at $B=3$ T. By converting the current $I$ to voltage $I/G_{xy}$, we see that $\Rds$ of both $\nu=2$ and $6$ oscillates only up to a voltage bias about 1 mV (see Fig.\,\ref{figcutv26}). Note that this range is approximately equal to the superconducting gap of MoRe ($\Delta_0 \approx 1.3$ meV). As this is the only energy scale at this order of magnitude in this system, the measurement further confirms that the underlying physics is due to superconducting correlations. 

\begin{figure}
\centering
\includegraphics[width=0.9\textwidth]{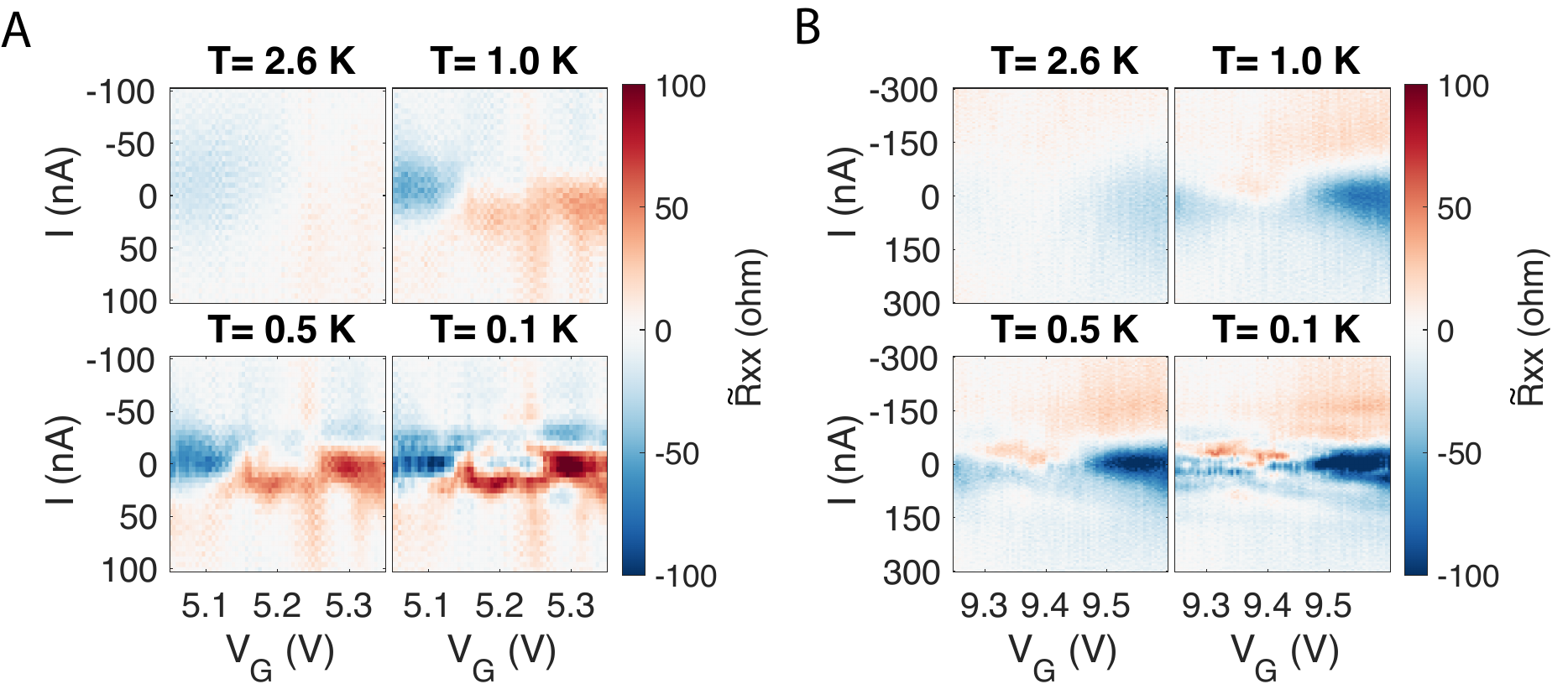}
\caption{
Temperature dependence of the bias-gate maps of $\Rds$ at $B=$ 3 T on the plateaus (\textbf{A}) $\nu=2$ and (\textbf{B}) $\nu=6$. }
\label{figmapT}
\end{figure}

This observation supports our interpretation in terms of interference among CAES. To further rule out any alternative explanations of these oscillations, we plot the bias-gate map of $\Rds$ on $\nu= 2$ and $6$ at various temperatures in Fig.\,\ref{figmapT}. The oscillation patterns gradually die off with rising temperature, leaving zero resistance independent of bias and gate voltage at 2.6 K. We also note that the bias-dependent oscillations for $\nu=6$ are more irregular (see e.g. Fig.\,\ref{figv6int}B) than those of $\nu=2$, suggesting the beating of multiple modes when the number of CAES is large.

\clearpage
\subsection*{S6. Tight-binding calculation}
\begin{figure}[htb]
\centering
\includegraphics[width=0.47\textwidth]{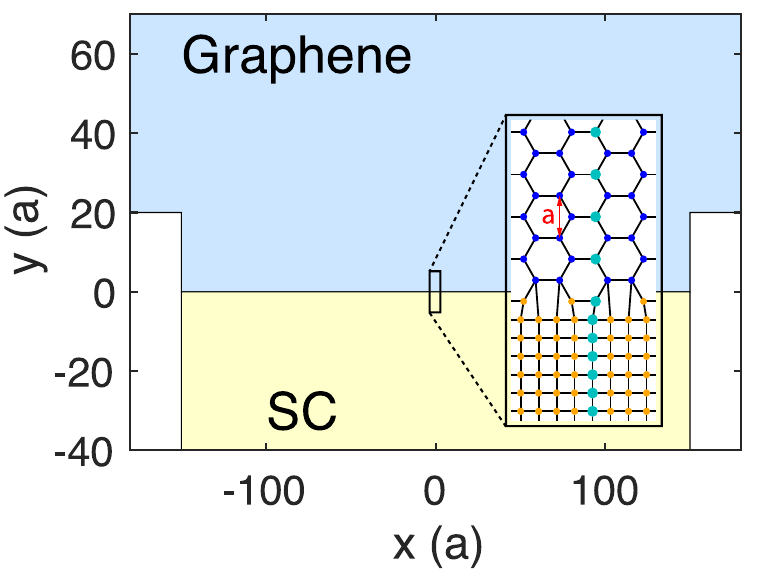}
\quad\quad
\includegraphics[width=0.44\textwidth]{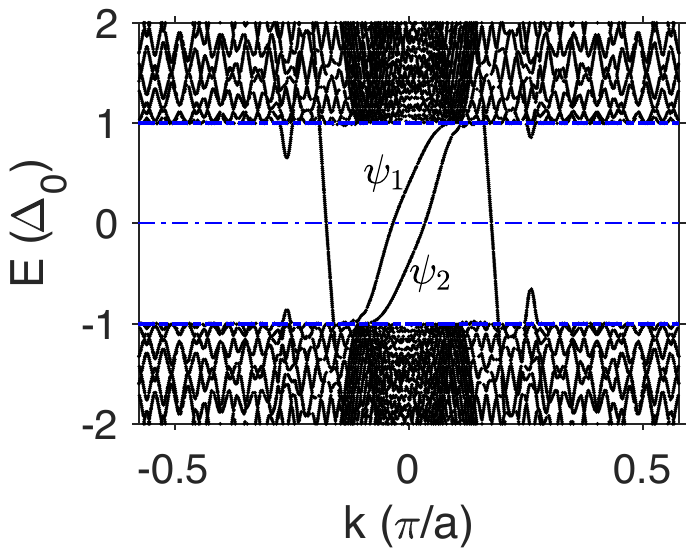}

\caption{
Schematic of the geometry (left panel): The graphene lattice (blue) is stitched to the superconductor lattice (yellow) as illustrated in the inset. The lattice parameter of graphene, $\mathrm{a}$, is marked red. The electron and hole density on the cyan atoms are plotted in Fig.\,\ref{fig:modedensity} for CAES $\psi_1$ and $\psi_2$.
Dispersion relation (right panel). Close to zero chemical potential ($E=0$), modes with a positive slope move along the quantum Hall-superconductor interface while those with a negative slope are located at the quantum Hall-vacuum edge and move in the opposite direction. The $\delta k$ between $\psi_1$ and $\psi_2$ gives rise to beating of the electron-hole hybridization, thus causing the resistance fluctuations seen experimentally.    
[$\mu_G=0.145\ t_G$, $B=0.0095\ h/e\mathrm{a}^2$.]
}
\label{fig:dispersion}
\end{figure}

In our numerical calculation, we consider an armchair graphene nanoribbon in contact with a superconductor, which is modeled as a square lattice with a superconducting gap. We avoid using a honeycomb lattice to represent the superconductor. In our earlier attempts to simulate the superconductor using the graphene lattice, the valley symmetry at the NS interface resulted in a pair of degenerate CAES. In this case the Andreev reflection probability is simply determined by the valley isospin of the edge states on the two sides of the superconducting contact \cite{S_akhmerov_detection_2007}. We emphasize that simulations using the square lattice for the superconductor are expected to represent a generic situation for a graphene-superconductor interface that does not have the valley symmetry. We have obtained similar results with both armchair and zigzag graphene nanoribbons. The graphene region is penetrated by a perpendicular magnetic field. We consider two basic geometries: 1) We calculate the transport properties using the setup shown in the left panel of Fig.\,\ref{fig:dispersion} with the lattice structure sketched in the inset. The square lattice of superconductor is stitched to the armchair edge of graphene lattice. The length of this interface is 301.5 $\mathrm{a}$, where $\mathrm{a}$ is the lattice parameter of graphene as indicated in the inset. Leads are attached to the left and right sides of the graphene region for calculating the transport properties. 2) To calculate properties in momentum space, such as the dispersion relations, we extend the interface shown in Fig.\,\ref{fig:dispersion} infinitely in the $x$ direction.

The graphene is modeled using the simple $\pi$-orbital nearest neighbor tight-binding model on a honeycomb lattice \cite{Ihnbook}. For simplicity, the model considered here is spinless. As a result, there is only one quantum Hall edge channel for the filling factor $\nu=2$ studied here.
The hopping energy in the graphene region and across the interface is set to be $t_G=3.033$ eV. The hopping energy in the superconductor region is $t_G/2$.
The magnetic field is incorporated with the standard Peierls substitution in a Landau gauge. The parameters are chosen such that the flux through a unit cell of the lattice is much less than a flux quantum,  $\Phi/\Phi_0 \!\ll\! 1$, even in the lowest Landau level (LLL, $\nu\!=\!2$). The density of electrons is typically different in the quantum Hall and superconductor regions, corresponding to a difference in the Fermi energy which we use as a crude model of the difference in work functions in the experimental system. The Fermi energy in the graphene region, $\mu_{G}$, is defined relative to the Dirac point, while in the superconductor region, $\mu_{SC}$ is defined relative to the band minimum. We set $\mu_{SC}=\mu_{G}+t_G$ to align the chemical potentials of the two regions.  

Superconductivity is simulated by using an electron and hole orbital on each site, coupled by a gap energy $\Delta_0=0.03\ t_G$. This is not, of course, an exact representation of Bardeen-Cooper-Schrieffer theory, but it does correctly mimic the effect of superconductivity on adjacent normal regions. As the transport quantities in which we are interested can be calculated entirely from the normal state properties, this suffices for our purposes. The Bogoliubov-de Gennes equation \cite{deGennesBook} is then solved assuming abrupt magnetic field and gap profiles, $B(y)=B H(y)$ and $\Delta(y)=\Delta_0 H(-y)$, where $H(y)$ is the Heaviside step function.

For non-interacting electrons, methods for calculating the interface modes, scattering wavefunctions, and transport are well known \cite{FerryGoodnickBook,DattaBook}. We use the package Kwant \cite{GrothKwantNJP14} to calculate the $S$-matrix, dispersion relations, and wavefunctions of the CAES along the quantum Hall-superconductor interface. From the $S$-matrix, we obtain the probability of an electron transmits through the CAES region as an electron versus a hole, $P_e-P_h$. This quantity is related to the experimentally measured differential resistance $\Rds$ by the Landauer-B\"uttiker formula obtained in section S3.

\begin{figure}[htb]
\centering
\includegraphics[width=0.49\textwidth]{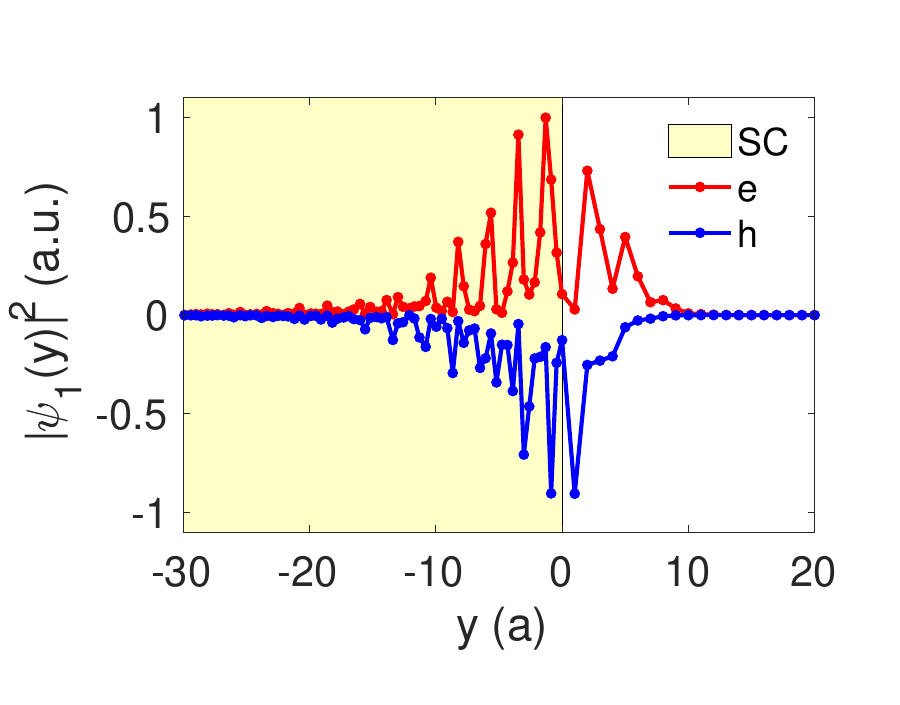}
\includegraphics[width=0.49\textwidth]{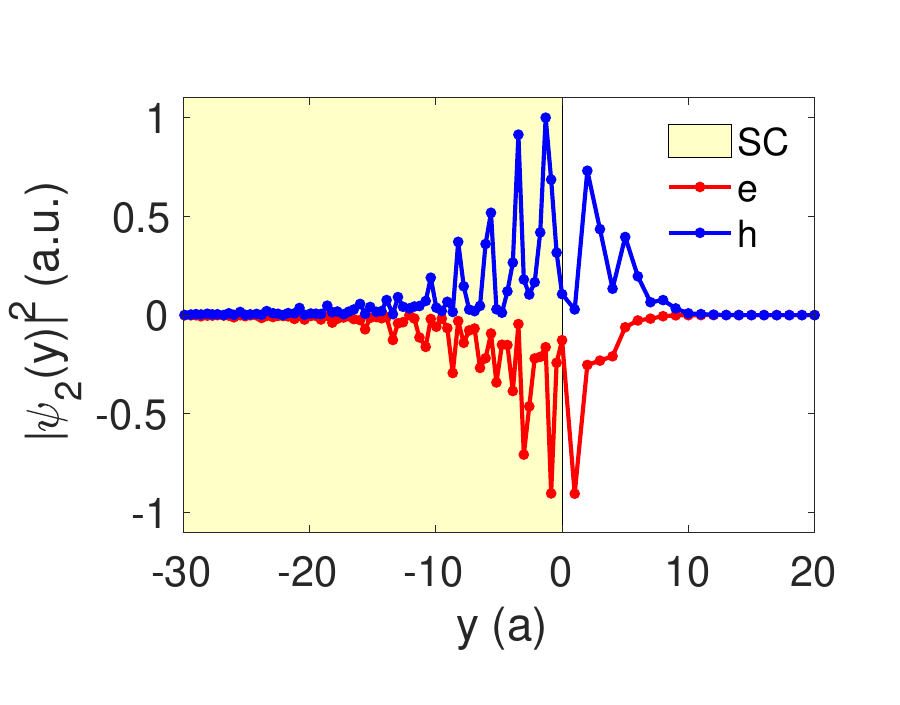}
\caption{
Wavefunctions of the interface modes at $E=0$ for $\nu=2$: electron and hole density (plotted with opposite sign for clarity) for the two CAES $\psi_1$ (left) and $\psi_2$ (right). Note that the electron and hole wavefunction in each mode are not the same and that the two modes are conjugates of each other. [$\mu_G=0.145\ t_G$, $B=0.0095\ h/e\mathrm{a}^2$.]
}
\label{fig:modedensity}
\end{figure}

We now turn to the results of our calculation. The dispersion relation of the interface modes, $E(k)$ where $k$ is the wavevector along the interface, is a key factor in explaining the experimentally observed oscillations. Thus, we show the dispersion relation for $\nu=2$ in the right panel of Fig.\,\ref{fig:dispersion}; for a zoom on the key interface modes, see Fig.\,3E of the main text. 
The edge modes that are propagating at the chemical potential ($E=0$) are clearly singled out---for $\nu=2$, there are two CAES along the quantum Hall-superconductor interface (positive slope) and two modes along the vacuum interface (negative slope). Spin degeneracy leads to a further doubling of the number of modes. Note that the group velocity ($dE/\hbar dk$) of the CAES is lower than that of the modes along the vacuum edge of graphene.
Because the wavevectors for the two CAES at the chemical potential are different, a superposition of such modes will undergo beating in position space. 

The electron-hole hybrid nature of CAES is immediately seen by looking at the corresponding transverse wavefunctions. The two CAES at the chemical potential are shown in Fig.\,\ref{fig:modedensity}. Note that the electron density of one mode is the same as the hole density of the other, as expected for a pair of modes that are charge conjugate. In a given wavefunction, electron and hole weights are approximately equal: in the case shown, the ratio of the integrated electron density to that of the hole is about $1.14$. However, as the electron and hole wavefunction of a given mode are certainly not the same, these modes are not charge neutral locally. 

In the calculation, a particularly striking illustration of the beating between the CAES is obtained by plotting the electron and hole probability densities for a single scattering wave, as shown in Fig.\,1D of the main text. In the case shown, an incoming electron (left vacuum edge) oscillates between electron and hole along the interface and then exits as a hole (right vacuum edge). The parameters used are $\mu_G=0.145\ t_G$, $B=0.0095\ h/e\mathrm{a}^2$ and $E=0$.

\begin{figure}[htb]
\centering
\includegraphics[width=0.49\textwidth]{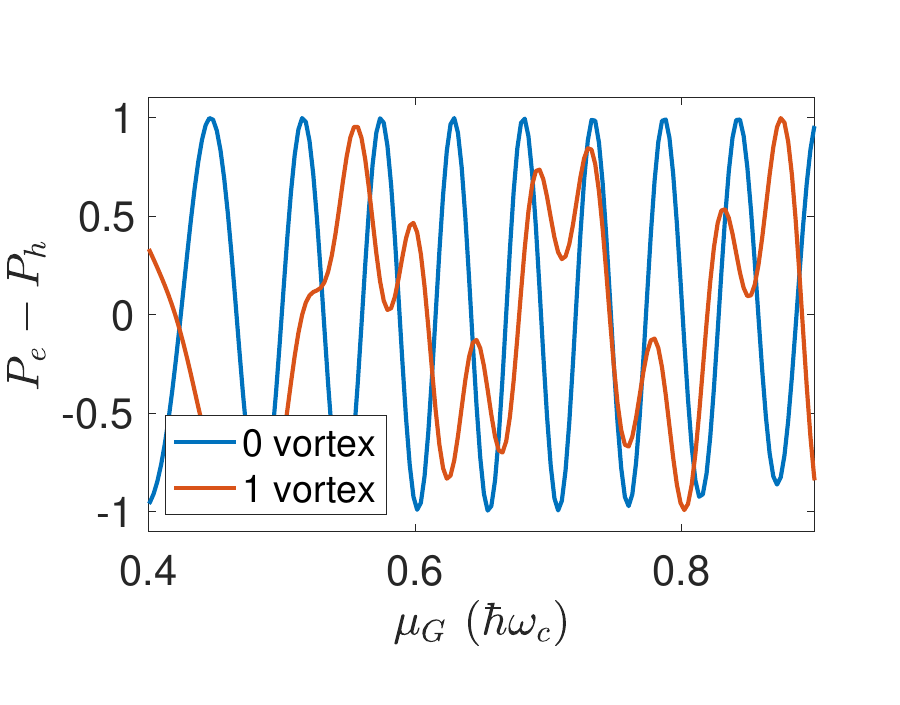}
\includegraphics[width=0.49\textwidth]{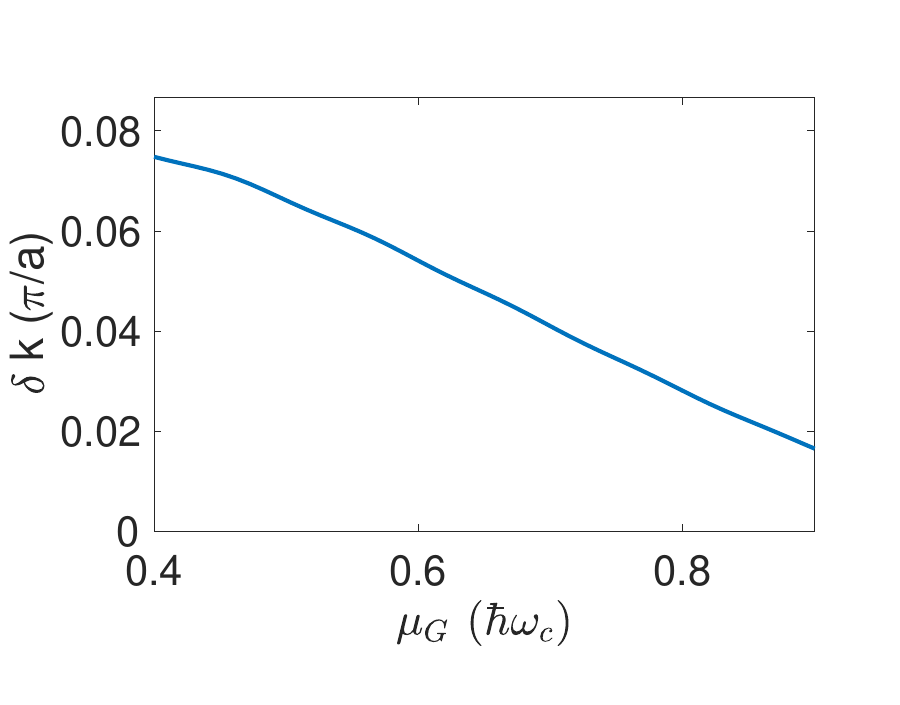}
\caption{ 
The probability difference $P_\text{e}-P_\text{h}$, as a function of the Fermi energy in graphene. When a twist in the phase of the superconducting order parameter $\Delta (x)$ is added to simulate the effect of a vortex in the superconductor, the transport changes drastically (blue for constant $\Delta$, red for twist). The difference in wavevector of the two modes is shown in the right panel.
[$E=0$, $B=0.0095\ h/e\mathrm{a}^2$. At this field the cyclotron energy $\hbar \omega_c=0.3\ t_G$.]
\label{fig:sim-GvsVg}
}
\end{figure}

Sensitivity of the observed transport quantities to \emph{gate voltage} is one of the main experimental signatures of interference effects among the chiral Andreev edge states. In the calculation, changing the gate voltage corresponds to changing the Fermi energy. The Fermi energy is changed uniformly throughout the system, in both the quantum Hall and superconductor portions. This has relatively small effect on the largely filled band of the superconductor, and mostly affects the graphene. Fig.\,\ref{fig:sim-GvsVg} shows calculation results for conversion at the quantum Hall-superconductor interface: the difference in reflection probability as an electron versus a hole would produce an oscillating $\Rds$ as per Eq.\,(\ref{eq:RDS}). On the right panel, the variation of the wavevector difference, $\delta k (\mu_G)$, that underlies the modulation is shown. 

\begin{figure}
\centering
\includegraphics[width=0.49\textwidth]{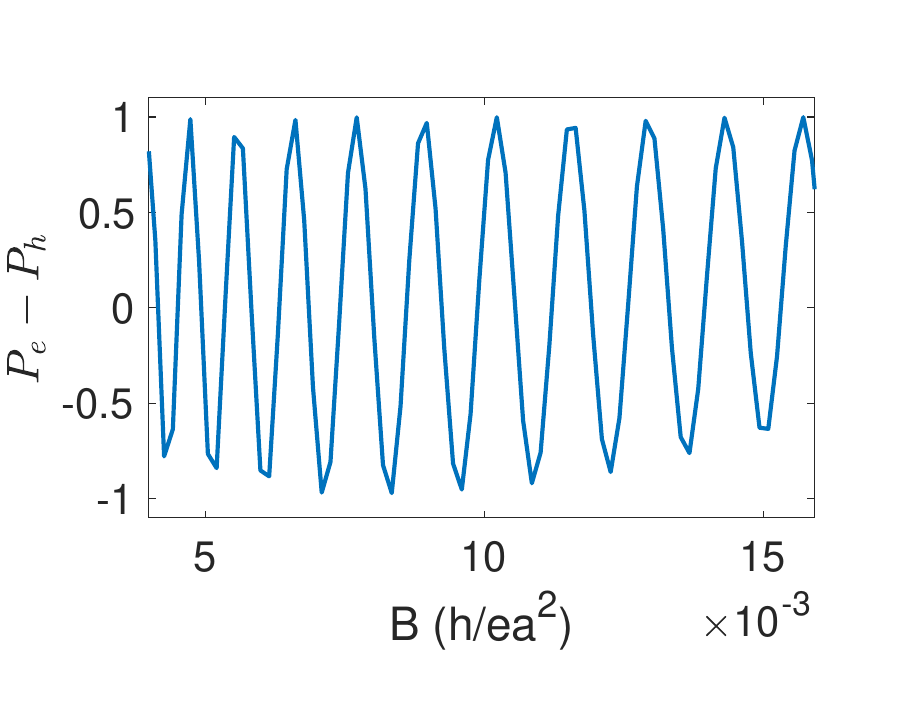}
\includegraphics[width=0.49\textwidth]{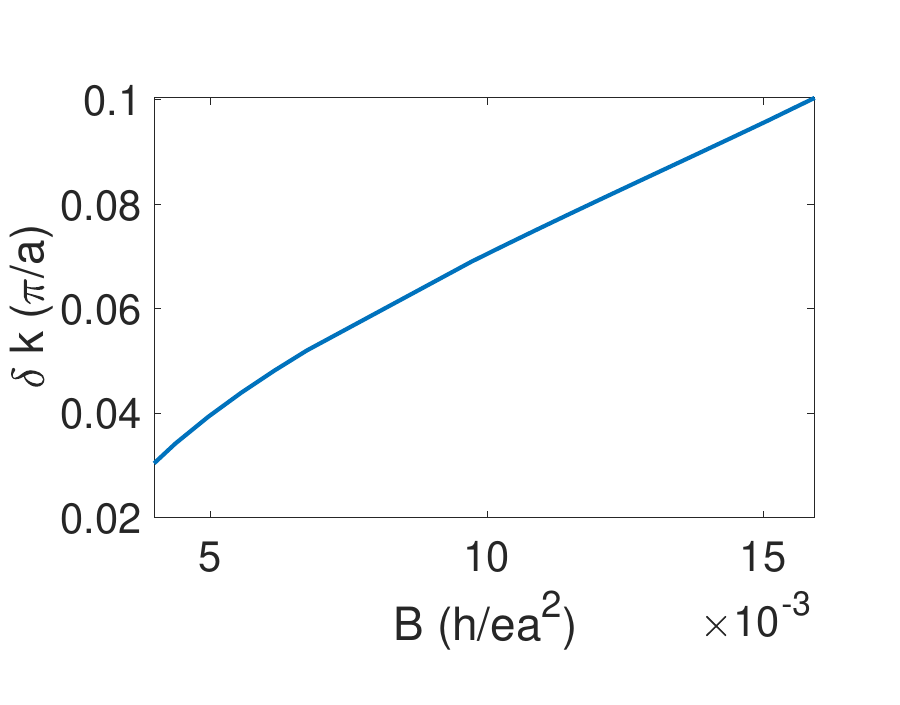}
\caption{
The probability that outgoing particle is an electron, $P_e$, minus that of a hole, $P_h$, as a function of the magnetic field B. These oscillations result from the variation of the wavevector difference between the two CAES, $\delta k$, as shown in the right panel.
[$\mu_G=0.145\ t_G$.] 
\label{fig:sim-GvsB}
}
\end{figure}

Changing the \emph{magnetic field} changes the phase of the wavefunctions and so modulates the beating. Fig.\,\ref{fig:sim-GvsB} shows the resulting change in the transmission probabilities $P_e$ and $P_h$ as a function of the magnetic field. The oscillation seen is the result of the change in $\delta k$, the difference between the wavevectors of the two hybrid modes, vs. $B$, as shown in the right panel.

To address the \emph{sharp switching} seen experimentally as a function of magnetic field, as in Fig.\,\ref{figS3}\&\ref{figvortex} and Fig.\,2 of the main text, we perform calculations with a complex gap $\Delta$ in which the phase varies along the quantum Hall-superconductor interface. We insert one vortex, which we crudely simulate by a twist in the phase by $\pi$ over a small distance. Fig.\,\ref{fig:sim-GvsVg} shows that this leads to a very different dependence of the electron/hole transmission probability on gate voltage, consistent with the experimental observations.

\begin{figure}
\centering
\includegraphics[width=0.5\textwidth]{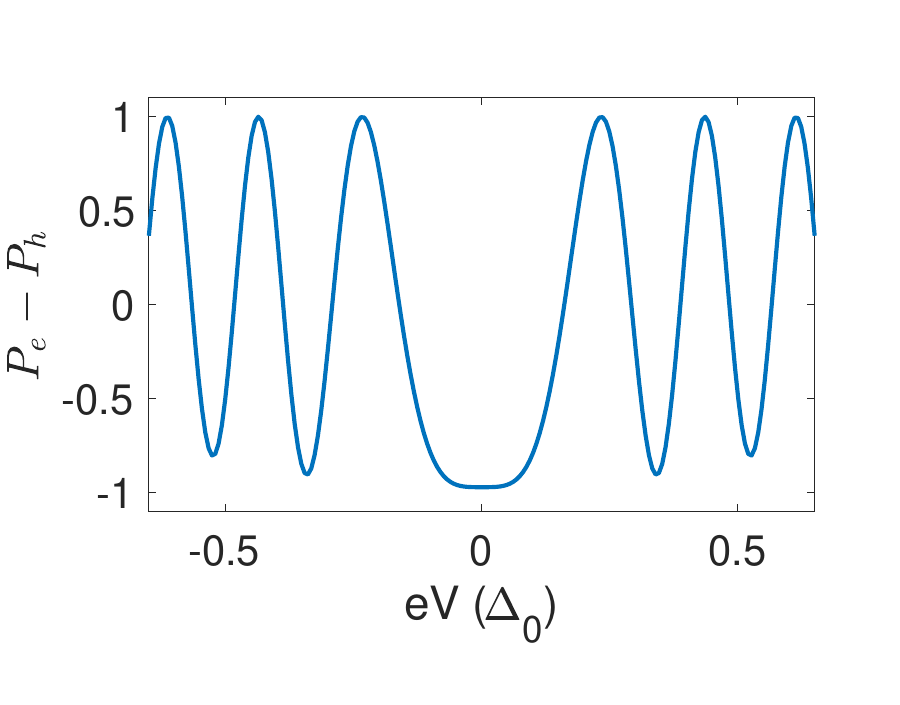}
\caption{
The probability difference $P_\text{e}-P_\text{h}$, as a function of the applied bias $E\!=\!eV$. Energy dependence of the reflection probabilities makes the differential resistance nonlinear, as seen experimentally. 
[$\mu_G=0.145\ t_G$, $B=0.0095\ h/e\mathrm{a}^2$.]}
\label{fig:sim-GvsMu}
\end{figure}

Variation of the transport properties with \emph{bias} comes from the energy dependence of $\delta k$ as one deviates from the chemical potential. Indeed, increasing the applied bias implies injecting higher energy electrons at the quantum Hall-superconductor interface. Introducing energy dependent reflection probabilities, $P_\text{e}(E)$ and $P_\text{h}(E)$, we show the probability of an incoming electron to exit the interface region as an electron versus a hole in Fig.\,\ref{fig:sim-GvsMu}. The energy dependence of the scattering produces oscillations in the differential transport similar to those observed experimentally. 
 
\subsection*{S7. Statistics of $\Rds$}

\begin{figure}
\centering
\includegraphics[width=1\textwidth]{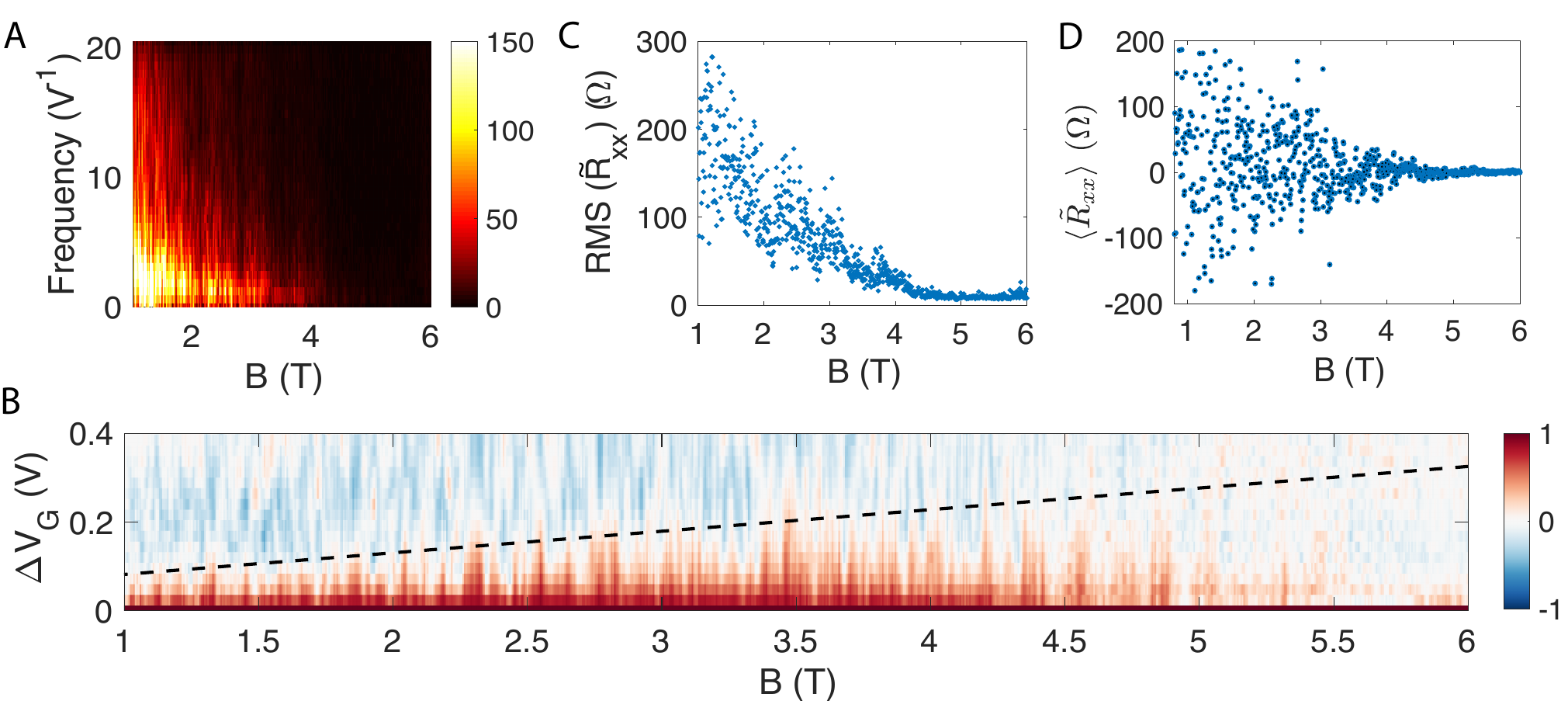}
\caption{
Statistical Analysis of $\Rds (V_G)$ on the $\nu=2$ plateau. The data used here are taken from Fig.\,2B. \textbf{(A)} $\left|FFT\left[\Rds (V_G)\right]\right|$ as a function of the magnetic field. The map is smoothed in the B direction by a moving average window of 20 mT. \textbf{(B)} The normalized autocorrelation function of $\Rds (V_G)$ as a funtion of magnetic field. The data are smoothed in the B direction by a moving average window of 20 mT. The dashed line shows 1/5 of the plateau width. \textbf{(C)} The root-mean-square (RMS) of $\Rds$ as a function of B. \textbf{(D)} The average value of $\Rds$ as a function of B.
}
\label{figFFT}
\end{figure}

We present the amplitude of the single-sided Fourier spectra of the downstream resistance ($\nu=2$), $\left|FFT\left[\Rds (V_G)\right]\right|$, as a function of magnetic field in Fig.\,\ref{figFFT}A. Unfortunately, the experimental signal oscillates only a few times over the width of the plateau (see e.g. Fig.\,1C). As a result, the most important information appears at the lowest frequencies, making it challenging to distinguish the relevant information from FFT window effects.

The dominant frequency of the oscillation is about 2 to 3 periods per volt, and the peak merges with a very broad background. This background corresponds to the random fluctuations of the signal, which are most likely explained by the disordered nature of the interface. The FFT amplitude at all frequency decreases with temperature or magnetic field, which indicates that all these components share the same origin.  

The autocorrelation of the signal (related to the Fourier spectrum) appears to be more informative. In Fig.\,\ref{figFFT}B we first present the normalized autocorrelation of $\Rds (V_G)$ for $\nu=2$ at various magnetic fields, extracted from Fig.\,2B. The autocorrelation is non-trivial for $1T<B<4T$, and gradually turns into noise above 5T. It appears that the region of positive autocorrelation grows proportionally with the width of the plateau. (The dashed line in Fig.\,\ref{figFFT}B shows one-fifth of the plateau width, which agrees well with the region of positive autocorrelation.) In addition to the Fourier spectrum, this observation suggests that $\Rds$ on average has 2 to 3 major oscillations on the  plateau, independent of the magnetic field. Finally, we calculate the standard deviation of $\Rds$ on the  plateau as a function of the magnetic field (see Fig.\,\ref{figFFT}C). The root-mean-square gradually reduces to the noise level above about 5 T.

For a single gate sweep measurement at a fixed field, we only observe a few major oscillations along the width of the plateau. As a result, a given trace may be preferentially weighted positively or negatively. However, when we measure  in a range of magnetic fields, we sample over various vortex configurations, and the result is on average neutral. In Fig.\,\ref{figFFT}D we averaged $\Rds$ over the gate voltage corresponding to a plateau, and presented $\langle\Rds\rangle$ vs. B.


\end{document}